\documentclass[article]{elsart}
\usepackage{graphicx,natbib}
\usepackage{amsmath}
\usepackage{amsfonts}
\usepackage{amsbsy}
\usepackage{amssymb}

\journal{Icarus}

\long\def\symbolfootnote[#1]#2{\begingroup%
\def\thefootnote{\fnsymbol{footnote}}\footnote[#1]{#2}\endgroup}

\newcommand{\x}{\ensuremath{\mathbf{x}}}
\renewcommand{\v}{\ensuremath{\mathbf{v}}}
\renewcommand{\u}{\ensuremath{\mathbf{u}}}
\renewcommand{\d}{\ensuremath{\partial}}
\newcommand{\w}{\ensuremath{\mathbf{w}}}
\newcommand{\g}{\ensuremath{\mathbf{g}}}

\newcommand{\kk}{\ensuremath{\mathbf{k}}}
\newcommand{\hh}{\ensuremath{\mathbf{h}}}
\newcommand{\jj}{\ensuremath{\mathbf{j}}}
\newcommand{\PP}{\ensuremath{\mathbf{p^{\text{c}}}}}
\newcommand{\WW}{\ensuremath{\mathbf{W}}}
\newcommand{\ani}{\ensuremath{\mathbf{a}}}

\newcommand{\eee}{\ensuremath{\mathbf{e}}}

\newcommand{\ey}{\ensuremath{\mathbf{e}_{y}}}
\newcommand{\ez}{\ensuremath{\mathbf{e}_{z}}}
\newcommand{\unit}{\ensuremath{\mathbf{1}}}
\newcommand{\FF}{\ensuremath{\text{FF}}}
\newcommand{\pib}{\mbox{\boldmath$\pi$}}

\begin{document}

\begin{frontmatter}

\title{Dense planetary rings and the viscous overstability}
\author[cam1,cam2]{Henrik N. Latter\corauthref{cor}},
\corauth[cor]{Corresponding author.}
\ead{henrik.latter@lra.ens.fr}
\author[cam2]{Gordon I. Ogilvie}
\ead{gio10@cam.ac.uk}
\address[cam1]{LRA, 
 \'{E}cole Normale Sup\'{e}rieure, 24 rue Lhomond, Paris 5, France}
\address[cam2]{DAMTP, University of Cambridge, Wilberforce Rd, Cambridge CB3
  0WA, United Kingdom}
\vskip3cm

\begin{abstract}

This paper examines 
  the onset of the viscous overstability in dense particulate
 rings. First, we formulate
 a dense gas kinetic theory that is applicable to the
  Saturnian system. Our model is essentially that
   of Araki and Tremaine (1986), which we show can be both simplified and generalised. 
    Second, we put
  this model
  to work computing the equilibrium properties of dense planetary
  rings, which we subsequently compare with the results of $N$-body
  simulations, namely those of Salo (1991). 
   Finally, we
  present the linear stability analyses of these equilibrium states, 
and derive criteria for the onset of
  viscous overstability in the self-gravitating and
  non-self-gravitating cases. These are framed
   in terms of particle size, orbital frequency, optical depth, and
  the parameters of the collision law. Our results compare favourably with the simulations of
  Salo \emph{et al.} (2001). The accuracy and practicality of the continuum model we
  develop
 encourages its general use in future investigations of nonlinear phenomena.

\end{abstract}

\end{frontmatter}

\newpage

\section{Introduction}

Of the many exciting spectacles delivered by \emph{Cassini}
images of complicated finescale structure in Saturn's B-ring are some of 
the most beautiful and intriguing (Porco \emph{et
  al.}, 2005). Irregular radial features on large scales, of some 10---100 km,
 have been associated with the B-ring
since \emph{Voyager}, but \emph{Cassini}
 has revealed that irregular structure is characteristic of
sub-kilometre
scales as well. In fact, a recent analysis of the UVIS stellar occultation data has uncovered
quasi-periodic variations on lengths as short as 150 m (Colwell \emph{et al.}, 2007).
It is has been suggested that these phenomena
 correspond to the saturated state of 
the viscous overstability, as examined by Schmit and Tscharnuter (1995,
1999). The viscous overstability is a linear instability that occurs
 when the viscous stresses vary with
density in such a way that energy from the
Keplerian shear can be directed into growing oscillations.
The most vigorously growing modes possess lengths of some 100---200 m
and could very well be responsible for irregular structure on 
sub-kilometre scales. On the other hand, 
it is perhaps unlikely that this mechanism generates the much longer
100 km features  first reported by
\emph{Voyager}, as had been originally hoped.
 These instead may be the responsibility of
ballistic transport (Durisen, 1995) or
electromagnetic effects (Goertz and Morfill, 1988).
Lending weight to the overstability hypothesis are
\emph{Cassini} ISS observations which report that
finescale structure favours regions in which the optical depth is
greater (Figs 5A and 5B, Porco \emph{et al.}, 2005); $N$-body simulations
show
the overstability is similarly sensitive to optical depth 
(Salo \emph{et al.}, 2001).
These new observations, and the theoretical challenges they throw up,
 encourage us to examine in greater detail when and why this instability occurs
(its linear theory) and the manner in which it saturates (its
nonlinear theory). 

 Researchers have mainly probed the 
onset of the instability with hydrodynamic models, sometimes allied with $N$-body
simulations (Schmit and Tscharnuter, 1995; Spahn \emph{et al.}, 2000;
Salo \emph{et al.}, 2001; Schmidt \emph{et al.}, 2001). In addition, Schmit and Tscharnuter (1999)
have conducted nonlinear hydrodynamical simulations and Schmidt and
Salo (2003) a weakly nonlinear analysis, both with isothermal models. 
The former study shows that the overstability takes the system into a disordered state in which power is
 transferred to longer and longer wavelengths, a process that only saturates
 if self-gravity is present.
 Schmit's and Tscharnuter's adoption of 
 reflecting boundary conditions, however, prohibits the development of
 the stable nonlinear travelling wave solutions that Schmidt
 and Salo (2003) discover. 
If dense regions of the B-ring indeed favour
 the nonlinear wave solution these probably possess a wavelength below the
resolution of \emph{Cassini}'s cameras but can be captured by
the UVIS or RSS instruments: in fact, it is very likely that the significant
150 m structures reported by the UVIS correspond to these
nonlinear waves.
Perhaps we can then intrepret the irregular structure
observed on slightly longer scales (1---10 km) as largescale modulations of the wave trains' amplitudes and/or
 wavenumbers, by analogy with water-waves and flame-fronts (Infeld
 and Rowlands, 1990), or alternatively they may 
 correspond to `defects' between wave trains of differing properties,
 by analogy with the dynamics of the complex Ginzberg-Landau
 equation (Aranson and Kramer, 2002). Certainly there is a lot of interesting work 
to be done examining the nonlinear aspects of this problem.

 In this paper, however, we shall reconsider the onset of the viscous overstability 
 in a dense ring. We employ a kinetic theory, and hence
 extend the dilute ring analysis of Latter and Ogilvie (2006).
 In that study it was argued that hydrodynamical models fail 
to capture important effects associated with
 the velocity anisotropy and non-Newtonian stress. Both we predict will play
 a part in the dynamics of a dense ring as well,
 a concern which motivates our adoption of the better-suited kinetic approach. 
Moreover, a kinetic model can be framed 
 in terms of parameters (particle size, collisional parameters, etc)
 that can be constrained by observation or experiment.  
 This parameterisation also makes a comparison with $N$-body simulations
 unproblematic. 
In contrast, hydrodynamics
 must make (sometimes speculative) prescriptions for the transport
 coefficients, because their functional dependence on the 
 observables is unclear. 
 
 The price one pays for using a kinetic model is mathematical complexity. 
 The dense gas collision terms are especially troublesome as they
 usually consist of complicated multi-dimensional integrals. Consequently, a number of
 assumptions are necessary to obtain some mathematical purchase on the problem.
 The model we employ is the one developed by Araki and Tremaine (1986)
 (herafter referred to as AT86),
 which assumes that the distribution function is a
 triaxial Gaussian.Such a distribution enabled Araki and Tremaine tor educe
 the collision integrals to four dimensions.
In this paper we show that further integrations
 are possible even when the coefficient of restitution is allowed to depend on
 normal impact velocity. In fact, if
 certain additional approximations are made, the remaining integrals can be executed
 by hand. These simplifications greatly facilitate the
 computation of the equilibrium states (previously a difficult
 numerical problem), and also open the way for more advanced
 analyses: with the elimination of these mathematical
 obstacles, a kinetic theory can be applied more generally,
 particularly to nonlinear behaviour that is presently beyond the
 range of direct $N$-body simulations.

In many ways, this work is similar to that of H\"{a}meen-Anttila who
 in a suite of papers developed the statistical mechanics of orbital
 elements using many of the same approximations and techniques 
(see H\"{a}meen-Anttila and Salo,
 1993, and references therein). The results derived using these methods should coincide with
 ours, though we find that working in a shearing sheet with a 
phase space of velocity and location elucidates local instabilities
 (such as the viscous overstability) much more clearly.

 \vskip0.3cm
 
 The format of the paper is as follows. The rest of this section
 introduces the viscous overstability and sketches
 out the
 basic issues pertinent to modelling a dense gas. In Section 2 we develop a
 dense ring model --- some of this section
 draws on previous work by Jenkins and Richman (1985) and AT86.
 Section 3 presents calculations and discussions 
 of dense ring equilibria plus a comparison with the $N$-body simulations of
 Wisdom and Tremaine (1988) and Salo (1991). In Section 4 we undertake
 a linear stability analysis of these states, thereby obtaining criteria for the onset of viscous overstability in the
 non-self-gravitating and self-gravitating cases. These results we compare with Salo \emph{et al.}\ (2001). In Section 5 we
 draw our conclusions and point out future work.

 \subsection{The viscous overstability}
 
 The viscous overstability is one of two instabilities of viscous origin that have
 been proposed to explain structure in planetary rings. The other, the
 viscous instability, occurs wherever the outward angular momentum flux
 is a decreasing function of surface density, a situation that initiates a clumping
 of matter into ringlets (Lin and Bodenheimer, 1981; Ward, 1981; Lukkari, 1981).
 For plausible parameters, the angular momentum flux of dense rings has been shown to be an \emph{increasing}
 function of surface density (Araki and Tremaine,
 1986; Wisdom and Tremaine, 1988), and so it is unlikely that
 this instability occurs in Saturn's dense rings. 

The viscous
 overstability, on the other hand, emerges from a Hopf bifurcation and, as the name suggests, can
 be characterised as an
overcompensation by the system's restoring forces: the
stress oscillation which accompanies the epicyclic response
in an inertial-acoustic wave will force the system back to
equilibrium so strongly that it will `overshoot'. Without self-gravity the longest
 lengthscales are the most susceptible to this runaway process, though they will grow slowly
 because the waves' growth rate is proportional to $k^2$, where $k$ is
 radial wavenumber. For sufficently small wavelengths, pressure
 extinguishes the instability and so there is a preferred intermediate
 scale upon which the viscous overstability grows most
 vigorously. When self-gravity is added the disk is more
 unstable, particularly on a band of intermediate
  wavelengths.

The mechanism of overstability relies
on: a) the
synchronisation of the viscous stress' oscillations with those of
density, and b) the viscous stress increasing sufficiently steeply
with surface density. In hydrodynamics only the latter
consideration is relevant, which furnishes the
criterion for overstability: 
$$\beta\equiv
(d\ln\nu/d\ln \sigma)>\beta^*,$$
 where $\nu$ is viscosity, $\sigma$ is surface density, and $\beta^*$ is a number dependent
on the thermal properties of the ring (Schmit and Tscharnuter,
1995). Generally, the viscous stress need not oscillate in phase
with the inertial-acoustic wave, in which event the
 transfer of energy between the two oscillations can be quite
 inefficient: usually the potential
 overstable modes are damped as a consequence. This is what happens in dilute
 particulate rings at small and intermediate collision frequencies
 (Latter and Ogilvie, 2006). The viscous stress of such rings, being
 predominately local (or `translational'), possesses a
 relaxation time of order the dynamical time scale, and so they lag
 behind the epicyclic variations. In contrast, the viscous stress in a dense
 ring
 is dominated by the nonlocal (`collisional') component
 which oscillates in phase with the inertial motion. Moreover, the effective viscosity profiles 
 computed by AT86's dense gas model and Wisdom
and Tremaine's (1988) particle simulations display a relatively steep increase
with optical depth (hence $\sigma$), suggesting that the effective $\beta$ of
the system may be sufficiently large to instigate overstability.

As a consequence, the linear behaviour of the viscous overstability has been
 thoroughly examined, mainly with hydrodynamics (in the papers mentioned
 earlier).
 However, it is upon
 the $N$-body study of Salo \emph{et al.} (2001) that we will
 concentrate. There 
the collisional evolution of
 $100$ cm radius particles were tracked in a shearing box situated in the B-ring undergoing
collisions that dissipate energy in accordance with the piecewise
 power law:
\begin{align} \label{bridge}
\varepsilon(v_{n}) = \begin{cases} (v_{n}/v_c)^{-p}, 
&\text{for}\quad v_n > v_c, \\
 1, & \text{for} \quad v_n \leq v_c,
\end{cases}
\end{align}
where $\varepsilon$ is the coefficient of restitution and $v_n$ is normal
impact speed. The parameters
$p$ and $v_c$ took the values deduced from the ice-collision
experiments of Bridges \emph{et al.}\ (1984), specifically: $p=0.234$
and $v_c=0.0077$ cm s$^{-1}$.
 These simulations show that a dense non-self-gravitating
disk is stable for all optical depths below about
four\footnote[1]{Throughout this paper `optical depth'
will denote normal geometric optical thickness.}. This critical value is
reduced to about one if
self-gravity's enhancement of the vertical restoring force is included.
 If self-gravity is fully modelled the emergence of the overstability is
 somewhat retarded by the vigorous percolation of
non-axisymmetric gravitational wakes, because these motions modify
 the effective viscous properties of the system.

It is problematic to obtain a general criterion for overstability from
simulations in terms of the various parameters, as this would require a
great number of individual runs.
 Moreover, the fluid dynamical criterion cannot be neatly related 
to the $N$-body results, the main problem being that the
quantities involved, namely $\nu$ and $\beta^*$, are not
straightforward functions of the $N$-body, or observable, parameters 
(particle size, orbital frequency, etc).
 Some of the hydrodynamic quantities can be numerically
obtained for a particular equilibrium state using simulations
 (Salo \emph{et al.}, 2001; Schmidt
\emph{et al.}, 2001), but the role of the various observable
 parameters is difficult to bring out, as these fluid dynamical quantities
 are
 tied to the particulars of a specific (and
lengthy) computation.
That said, some of the appurtenant results so obtained
(growth/decay rates mainly) are in good agreement with the
simulations, even if the assumptions implicit in a
 hydrodynamic model possibly lead to errors. 

 We seek to use kinetic theory to remedy this problem, by
 providing a criterion for the onset of overstability, specifically an
 estimate of the critical optical depth $\tau_c$, as a
 function of the kinetic parameters of the system, which connect to the observable data:
 particle radius $a$, particle mass $m$, 
local orbital frequency $\Omega$, and the parameters that appear in the elasticity
 law. 
 
 \subsection{Dense gases}

  It was not until the late 1980s that it was fully appreciated that
 nonlocal `dense' effects were critical and widespread in Saturn's
 rings. Perhaps this shift was initiated most by the
 theoretical papers of AT86 and Araki (1988, 1991), and
 the simulations of Wisdom and Tremaine (1988) and Salo (1991). In
 fact,
 the issue had been presaged by Brahic (1977) and examined
 in some detail
 by H\"{a}meen-Anttila (1982) and Shukhman (1984).
Essentially, Araki and
Tremaine and Wisdom and Tremaine revealed that (a)
collisional contributions are crucial to the equilibrium
when collisions are sufficiently inelastic and (b)
 the experimentally derived elasticity laws of
\eqref{bridge} predict that collisions are dissipative enough for this
`cold' regime to be ubiquitous in Saturn's rings, given appropriate
particle sizes (Marouf \emph{et al.}, 1983; Zebker
\emph{et al.}, 1985).
In addition, both simulations and theoretical models show that the
dynamics of small particles (for which nonlocal effects are less
important) strongly couples to the dynamics of the largest
particles (for which nonlocal effects are important) both in
two-size systems and in
polydisperse disks exhibiting power-law size
distributions akin to Saturn's (Stewart \emph{et al.}, 1984; Lukkari, 1989; Salo,
1987, 1991, 1992b).

 To capture theoretically this dense regime we turn to the Enskog model which revises the
 Boltzmann theory for moderately dense gases
(Chapman and Cowling, 1970).
 The Enskog formalism distinguishes two additional processes not captured by
 Boltzmann kinetics;
the first is associated with a large filling factor, where filling factor (\FF) denotes the proportion of 
space occupied by particles and is defined, for spheres, through
$$\FF\equiv \tfrac{4}{3}\pi n a^3, $$
where $n$ is volumetric number density and $a$ is particle radius. The second process
 is associated with the
  collisional transfer of particle properties such as
momentum and energy. We describe these in turn.
 
When particles take up a significant portion
of space (i.e.\ when $\FF$ is not small) the volume in which they may
move is reduced, and possible colliders may be screened by other
particles (Chapman and Cowling, 1970). This means that the statistics of two impacting
particles must include the influence of their neighbours and, as a consequence,
the evaluation of
the collision
frequency must take space
correlations into account.
Overall, this leads to an enhancement of the collision frequency
$\omega_c$ which, in the Enskog theory, is approximately quantified by
a factor $Y(\FF)$ (the `Enskog factor'). This
cannot be calculated within the bounds of kinetic theory but must be
determined separately, usually from data gathered in molecular dynamics
experiments (Araki and Tremaine, 1986; Jenkins and Richman, 1984;
P\"{o}schel and Brilliantov, 2004). Inelastic and spinning
systems may differ in this respect from their elastic and spinning
counterparts because they can develop postcollisional correlations in their
particles' tangential velocity. 
Inelastic collisions diminish the normal component of
the relative velocity but conserve the tangential component, and as a
result subsequent collisions lead to a statistical alignment of
neighbouring particles' traces (Brilliantov and P\"{o}schel, 2004). Other than possibly forming
vortices, this effect can reduce the local value of the collision
frequency, and so the Enskog factor of an inelastic ensemble may be
smaller than that of a corresponding elastic system (see P\"oschel,
Brilliantov and Schwager, 2002). That all said, in ring applications
the net effect seems negligible.

There are two ways properties may be transferred in a particle gas: their free carriage by
particles \emph{between} collisions and their transmission from one particle to
another \emph{during} a collision. In a dilute gas, the former so-called
`local', or translational, mode dominates the
latter because particles travel
relatively long distances between collisions. In a dense gas, the mean
free path is much reduced and can approach a particle radius, in which case the
finite size of the particles is sufficiently large for the
exchange of properties between colliding particles to compete with, or even
dominate, translational transport.
Estimates for the relative magnitude of the two types are easy to derive. Consider the
transport of momentum across a plane by a gas characterised by mass density
$\rho$ and velocity
dispersion $c$, and composed of particles of radius $a$. 
The magnitude of the momentum flux density
due to the free carriage of particles is
 of order $\rho c^2$. On the other hand, the flux density of momentum carried
 in collisions
from the
centres of all the particles on one side of the surface to all the
particles on the other side is of order
$(\rho\, c)a\,\omega_c$. This implies the scaling:
\begin{equation} \label{nlest1}
\frac{\|\mathbf{p}^{\text{c}}\|}{\|\mathbf{p}^{\text{t}}\|}\sim \frac{a\,
  \omega_c}{c}=(\omega_c/\Omega)R,
\end{equation}
where $\mathbf{p}^{\text{c}}$ and $\mathbf{p}^{\text{t}}$ refer
to the collisional and translational pressure tensors respectively,
$\Omega$ is the local orbital frequency, and $R$ is defined by
\begin{equation} \label{RRR}
R\equiv \frac{a \Omega}{c}.
\end{equation}
Elsewhere $R$ is called the Savage and Jeffrey $R$-parameter (Savage
and Jeffrey, 1981; Araki, 1991), and it quantifies the ratio of shear
motions to velocity dispersion. Here $\Omega$ is a substitute for
shear rate. Note that it is $(\omega_c/\Omega) R$ that quantifies the importance of
collisional transport in the ring \emph{dynamics}. In contrast, an analogous argument
shows that $R^2$ quantifies the importance of collisional effects in the
ring \emph{energetics}: because heating and cooling are proportional to
collision frequency the $\omega_c/\Omega$ can be dropped from the estimate.

The effects of large filling factor and collisional transport usually
work in tandem, though for very high or very low optical depths there
are cases when one can exist without the other. This is expressed in the scaling:
\begin{equation} \label{nlest3}
\tau\sim \FF/R.
\end{equation}
For a substantial discussion on this subject see Araki (1991).
In
Saturn's rings $R$ is probably of order unity (Salo, 1992b), which means that filling factor
effects are as important as the optical depth is
large. In low optical depth regions, such as the C and
 D-rings and the Cassini division, filling factor effects are hence
negligible. In contrast, collisional transport/production effects will be important
throughout the rings on account of their low velocity dispersion
($R\sim 1$).

\vskip0.3cm

Kinetic models and $N$-body simulations, in order to make any progress, must introduce a number
of simplifying assumptions. Typically, ring particles are taken to be
identical, perfectly smooth, hard, indestructible spheres which
dissipate energy in collisions according to a simple `elasticity law'
such as \eqref{bridge} (the `billiard ball' model). Though these presumptions
 facilitate calculations,
 it is unclear how appropriate some of these are to the real
rings of Saturn. Neglecting spin and size-distribution should give
qualitatively correct results for those particles which
dominate the dynamics (near 100 cm size). Also it is plausible that
 the timescale of the size
dynamics (erosion and accretion, etc) may be much larger than the
dynamical timescale ($\Omega^{-1}$), meaning these processes do not interfere
with the instabilities we study. In any case, the modelling of spin is hampered by the
fact that the frictional properties of the ice particles are poorly
constrained.
We admit that `smoothing' over the complicated details of an actual collision
with a simple $\varepsilon$ law is a bold approximation, particularly 
as there is considerable
 uncertainty about the relevant physics
for real ring particles. A number of terrestrial 
experiments, however, have been conducted with solid ice spheres (Bridges \emph{et al.}, 1984; Hatzes
\emph{et al.}, 1988; McDonald \emph{et al.}, 1989; Hatzes \emph{et al.}, 1991; Supulver \emph{et al.}, 1995; Supulver \emph{et
  al.}, 1997). These show that the coefficient of restitution varies considerably
as the physical condition of the contact surface is frosty, sintered, 
or sublimated. Depending on the surface condition, 
various processes ensue, such as erosion, sticking, mass
transfer, and regolith compactification, all of which influence
$\varepsilon$. But it has also been suggested that a particle may resemble more of a rubble
pile (a `dynamic ephemeral body') held together by mutual gravitation
(Weidenschilling \emph{ et al.}, 1984) and/or adhesion (Albers and
Spahn, 2006), in which case the collisional
dynamics will be very different --- and the assumption of `hardness',
and a neat $\varepsilon$ law itself, perhaps unsupportable. Unfortunately, 
we lack detailed information about the physical state of a ring
particle, and given this uncertainty assume that laws
such as \eqref{bridge} capture the correct qualitative behaviour,
 though the parameters which appear
in them may be subject to much variability.

\section{The kinetic model}

\subsection{Governing equations}

 Consider a gas of indestructible identical nonspinning inelastic spheres of
mass $m$ and radius $a$, and with phase space distribution $f(\x,\v, t)$. The number
of particles located in the volume $d\x$ centred at $\x$ with
velocities in the range $d\v$ centred at $\v$ at time $t$ is defined as
$  f(\x,\v,t)d\x\, d\v$. 
Number density $n$, bulk velocity $\u$, and the pressure tensor
$p_{ij}$ are defined, consequently, through
\begin{align} \label{ch2.1}
n(\x,t) &\equiv \int  f\, d \v, \\
n u_i(\x,t) &\equiv \int  v_i f\, d \v, \label{ch2.2}\\
 p_{ij}(\x,t) &\equiv \int (v_i-u_i)(v_j-u_j)\, f d \v. \label{ch2.3}
\end{align}   
 The phase space distribution 
satisfies a kinetic equation which is distinguished by a collision
operator, $(\d f/\d t)_c$, whose precise form reflects the collisional
microphysics. This equation is presented after we describe the geometry of the
problem.

The particle gas inhabits the shearing sheet (see Goldreich and Lynden-Bell, 1965).
 This is a convenient 
representation of a
differentially rotating disk in which a small patch, centred on
a point moving on a circular orbit at
$r=r_{0}$, is represented as
a sheet in uniform rotation, $\Omega(r_0)\, \ez$, and subject to a
linear shear flow, $\u_{0}=-2A_{0}x\, \ey$. The local rectilinear
coordinates $x$ and $y$ point in the radial and azimuthal directions
respectively and
$A_{0}=-\frac{1}{2}r_{0}(d\Omega/d r)_0$. A Keplerian
disk requires $A_0=\frac{3}{4}\Omega_0$, where
$\Omega_0=\Omega(r_0)$. 
This approximation is valid
if the radial wavenumber of typical variations $k$
 satisfies $|kr_0| \gg 1 $, which is certainly the case for the phenomenon of
 interest. From now the subscript `$0$' will be dropped.

In terms of the peculiar velocity, $\w\equiv\v-\u$, the kinetic equation
for particles in the shearing sheet is
\begin{equation} \label{E3}
\frac{\d f}{\d t} + (w_{i}+u_{i}) \frac{\d f}{\d x_{i}} 
-\left[ \frac{\d u_{i}}{\d t} + (w_{j}+u_{j})\frac{\d u_{i}}{\d
    x_{j}} - F_{i} \right] \frac{ \d f}{\d w_{i}} = \left( \frac{ \d f}{\d t} \right)_{c},
\end{equation} 
where the force per unit mass is 
$$ F_{i}= -\frac{ \d (\phi_P+\phi_D)}{ \d x_{i}} - 2 \epsilon_{ijk} \Omega_{j}
  (w_{k}+u_{k}). $$
The appropriate centrifugal-gravitational potential of
  the planet is denoted by $\phi_P$ and
  the disk's gravitational potential by $\phi_D$. The tensor $\epsilon_{ijk}$ 
is the alternating tensor and the
  angular velocity is $\mathbf{\Omega}=
  \Omega\, \ez$. 
By multiplying \eqref{E3} by $1$, $ w_{i}$ and $
  w_{i}w_{j}$ and then integrating over all $\w$ we derive the continuity equation,
\begin{equation} \label{E4}
\d_{t} n + \d_{k} (n u_{k}) = 0,
\end{equation}
the equation of motion,   
\begin{equation} \label{E5}
n\left( \d_{t} u_{i} + u_{k} \d_{k} u_{i} \right) = -2n
\epsilon_{ijk} \Omega_{j} u_{k} - n \d_{i} (\phi_P+\phi_D) - \d_{j}p_{ij}+ m_{i},
\end{equation}
and the pressure tensor equation,
\begin{align} \label{E6}
\d_{t} p_{ij} + u_{k}\d_{k}p_{ij} = -p_{ik}\d_{k}u_{j}
&-p_{jk}\d_{k}u_{i} -p_{ij}\d_{k}u_{k}
-2\epsilon_{ikl}\Omega_{k}p_{lj} \notag \\ & \hskip2cm -2\epsilon_{jkl}\Omega_{k}p_{li}  
-\d_{k} p_{ijk} + q_{ij}, 
\end{align}
where $p_{ijk}$ is the third-order moment,
\begin{equation}
p_{ijk} = \int  w_{i} w_{j} w_{k}  f d \mathbf{w}, 
\end{equation}
 the collisional change in the first moment is
\begin{equation} 
m_{i}= \int w_i \left(\frac{\d f}{\d t}\right)_c d \mathbf{w}, 
\end{equation}
 and the collisional change in the
 second moment is
\begin{equation} \label{qqq}
q_{ij }= \int w_i w_j \left(\frac{\d f}{\d t}\right)_c d
\mathbf{w}. 
\end{equation}
For notational brevity we have set $ \d_{i} = \d/\d x_{i}$.
If nonlocal transport effects are neglected then $m_i$ is zero on account 
of momentum conservation. Jenkins and Richman (1985) show how to rewrite this term as a flux, so that
$$m_{i} = -\d_j p_{ij}^{\text{c}}, $$ where
$p_{ij}^{\text{c}}$ is the collisional pressure tensor. We will work
with this form for the rest of this paper.


\vskip0.3cm
The moment equations are vertically integrated, 
 a task that is easily performed once we assume that
 $u_x$, $u_y$, and $p_{ij}$ are independent of $z$.
 Thus the disk is assumed to be vertically isothermal. 
In addition, vertical averaging permits us to close the hierarchy of the momemt equations.
In the pressure tensor equation Eq.~\eqref{E6}, 
this procedure removes the dominant $z$ derivatives of the
 third-order moment, $\d_z p_{ijk}$, leaving only its
  horizontal derivatives, which scale relative to the
 other terms like $H k$, where $H$ is disk semithickness and $k$ is the
 wavenumber of a typical horizontal variation. We expect these variations to be much
 longer than the disk thickness and so treat these flux terms as negligible.
 In summary, 
  the closed governing equations we derive
 validly describe behaviour on scales satisfying $H\ll\lambda\ll r_0$, where $\lambda$ is
 wavelength. 
 
We define optical depth as $\tau\equiv \pi a^2 N$, where $N$ is surface
number density. We also introduce the velocity dispersion tensor:
$$ W_{ij}\equiv p_{ij}/n.$$
This then allows the vertically-averaged equations
 to be recast as
\begin{align} \label{Ebeast}
D_t \tau &= -\tau \d_\alpha u_\alpha, \\
D_t u_\alpha &= -\d_\alpha (\Phi_P+\Phi_D) - 2\epsilon_{\alpha z \beta} \Omega
\,u_\beta -\frac{1}{\tau}\,\d_\beta\left[\tau\,W_{\alpha
    \beta}+ \tau\,P^\text{c}_{\alpha \beta}/N \right], \\
D_t W_{ij} &= -W_{jk} \d_k u_i - W_{ik} \d_k u_j - 2 \epsilon_{i z
  k}\Omega\, W_{kj} - 2 \epsilon_{j z k} \Omega\, W_{ki} +
Q_{ij}/N, \label{E10}
\end{align}
where $D_{t}= \d_{t}+ u_{\gamma}\d_{\gamma}$, Greek indices run
 only from $x$ to $y$ and upper case denotes vertical
 integration except for $\Phi_P$ and $\Phi_D$ which are vertically
 averaged (if needed).
 We
 approximate the central body as perfectly spherical, which accounts
 for the former potential, as then
$ \Phi_P= 3\Omega^2\, x^2/2 +\mathcal{O}(x/r_0).$ This is acceptable
because the precessional effects associated with planetary oblateness
 are unimportant for the instabilities we study.
The latter potential $\Phi_D$ must be obtained from Poisson's
 equation:
\begin{equation} \label{Poisson}
\nabla^2 \phi_D = 4 \pi m G n,
\end{equation}
 where $G$ is the gravitation constant

Finally, we are left with the vertical component of the equation of
motion from which it is possible to extract an equation for the mean
vertical displacement of the disk $Z$ (see Shu and Stewart, 1985).
 In this paper we do not investigate 
vertical warping of the disk. We assume vertical hydrostatic
equilibrium and symmetry
about the plane $z=0$; thus $Z=P_{xz}=P_{yz}=0$. The
vertical balance furnishes an equation for the vertical stratification of density
\begin{equation} \label{verteqm}
-n\Omega^2\,z-4\pi G\,n\,m\int_0^z\,n(\hat{z})d\hat{z}\,-\d_z\left[
 nW_{zz}+P^{\text{c}}_{zz}\right]=0.
\end{equation}
This integro-differential
equation for $n$ can be manipulated into a second order ODE, which
must be solved concurrently with Eqs
\eqref{Ebeast}--\eqref{E10}. Alternatively, an ansatz for the form of 
$n$ can be employed, such as a Gaussian (AT86) or a
polytrope (Borderies, Goldreich and Tremaine, 1985; Mosquiera, 1996),
 in which case \eqref{verteqm} becomes an algebraic
equation for either the scale height or semithickness of the disk, $H$.

The assumption of vertical hydrostatic equilibrium, though convenient,
may be inappropriate when dealing with many dynamical
phenomena, including overstable oscillations and density waves. This
is because the
timescale upon which vertical equilibrium is established should be
of the same order as the dynamical time. It is possible to
derive a dynamical equation for $H$, which could capture this
effect. But for simplicity we persist with the hydrostatic model in
this paper.

\subsection{Construction of the collision integrals}

In the dilute disk analysis of Latter and Ogilvie (2006) the microphysical
details of the
collisional interactions could be largely sidestepped by adopting a BGK
model (Shu and Stewart, 1985; Bhatnagar, Gross and Krook, 1954).
 The collision terms in this case were simple, analytic, and
local in space. Regrettably, no comparable approximation exists for the
nonlocal collisional interactions of a dense gas. Hence we wade into
the mathematics of the Boltzmann-Enskog collision theory
and build the terms $q_{ij}$ and $p_{ij}^{\text{c}}$ from first
principles.
In their construction it will be assumed that particles only interact
with each other
through collisions and not also through gravity. This omission eliminates
the possibility of  gravitational focusing and scattering.

\subsubsection{The kinematics of binary collisions}

In order to ascertain the consequences of particle collisions collectively we must
establish the basic physics of a single collision first.
Consider two colliding particles of mass $m$ and radius $a$
travelling with velocities $\v_{1}$ and $\v_{2}$ immediately before the collision
and $\v_{1}'$ and $\v_{2}'$ immediately afterwards. The relative velocities pre
and postcollision are $\g=\v_1-\v_2$ and
$\g'=\v_1'-\v_2'$, and their locations at the moment of collision are
$\x_1$ and $\x_2$. The collision is assumed to be inelastic
and so the component of $\g$ normal to the
impact is reduced by the factor $\varepsilon$, the coefficient of
restitution. Hence we write
$$ \g'\cdot\kk = - \varepsilon\, \g\cdot \kk $$
where $\kk$ is the unit vector directed from the centre of particle 1
to that of particle 2 so that
$$ \x_2-\x_1=2a\,\kk.$$
Obviously, $\g\cdot\kk>0$ (the impact
velocity cannot be negative). Also, it is assumed
that $\varepsilon$ is a function of normal impact velocity, $v_n\equiv \g\cdot\kk$.

This prescription is sufficient to set the postcollisional velocities:
\begin{equation}
\v_1'=\v_1-\jj, \qquad \qquad \v_2'=\v_2+\jj,
\end{equation}
where 
$$ \jj=\frac{1}{2}(1+\varepsilon)(\g\cdot\kk)\kk,$$
and
$m\jj$ is the momentum transferred from particle 1 to particle
2.

The velocity of a particle may be expressed as the sum of the
local mean velocity $\u(\x,t)$ and the peculiar velocity of the
particle, $\w= \v-\u$. The precollisional relative velocity is therefore
$$ \g=\w_1-\w_2+\hh,$$
where $\hh$ denotes the difference in mean
velocity between the locations of the particles: thus
$\hh\equiv\u(\x_1,t)-\u(\x_2,t)$. Since the collision is instantaneous the
mean velocity does not change in the collision, and
 because the particles are assumed perfectly smooth, the tangential component of $\g$ is conserved.

 The prescription above
determines the total change in the peculiar velocity dyadic
 in a binary collision,
\begin{align} 
\Delta(\w\w) &= \w_1'\w_1'+\w_2'\w_2' - \w_1\w_1 - \w_2\w_2 \notag \\
& = \frac{1}{2}(1+\varepsilon)(\g\cdot\kk)[
(\hh-\g)\kk+\kk(\hh-\g)+(1+\varepsilon)(\g\cdot\kk)\kk\kk ]. \label{2ndm}
\end{align}
Hence the total change in peculiar kinetic energy is
\begin{equation} \label{enervate}
\frac{1}{2}\,m\Delta(w^2)=-\frac{1}{4}\,m(1-\varepsilon^2)(\g\cdot \kk)^2+m\,\mathbf{j}\cdot\hh.
\end{equation}
The first term in \eqref{enervate} represents the dissipative loss of
energy (zero when $\varepsilon=1$). The second term represents the
gain in energy from the transfer of momentum across the mean shear flow. If
the particle size is sufficiently small so that
$|\hh|\ll|\w_1-\w_2|$ (which corresponds loosely to $R\ll 1$, see Eq.~\eqref{RRR}), then this
contribution can be neglected.

\subsubsection{The statistics of binary collisions}

We now determine the frequency of a particular collision specified by
 $\v_1$, $\v_2$, $\x_1$, and $\x_2$. This
requires the introduction of 
 the pair distribution $f^{(2)}$ which is defined so that
$$ f^{(2)}(\v_{1},\x_{1},\v_{2},\x_{2},t) d\v_1 d\v_2 d\x_1 d\x_2 $$
is the probable number of particle pairs located in the small
volumes $d\x_1$ and $d\x_2$ centred at $\x_1$ and $\x_2$, with
velocities in the volumes $d\v_1$ and $d\v_2$ centred at $\v_1$ and
$\v_2$ in velocity space, at a time $t$.
For a dilute gas, in which particles are well spaced and travel
relatively long distances between collisions, we may appeal to
the assumption
of molecular chaos (\emph{Stosszahlansatz}) in order to simplify the
distribution.
 This assumption states that
there exists no correlation between the velocities and positions of
any two particles before their collision.
 Consequently, we write $f^{(2)}(\v_{1},\x_{1},\v_{2},\x_{2},t)$ as the product of
two single particle distributions, $f(\x_1,\v_1,t)$ and
$f(\x_2,\v_2,t)$. But in a dense gas the probable
positions of two colliding particles will be influenced by
the presence of their neighbours. As mentioned earlier, we may
 approximate this by including the Enskog factor $Y$,
which we take as a function of filling
factor, and hence number density, $n(\x)$. This last assumption
limits $Y$ to express position correlations only and 
 is justified for elastic gases in which number density is 
nearly uniform. As we only explore homogeneous equilibrium or small
deviations from it, this aspect of the assumption is acceptable for our purposes (AT86). More generally, velocity or higher order
correlations may need to be taken into account.
 In summary, we write for two identical particles of radius $a$ in contact
\begin{equation} \label{Ensk}
f^{(2)}(\v_{1},\x_{1},\v_{2},\x_{1}+2a\kk,t)=
Y[n(\x_1+a\kk)]f(\v_1,\x_1,t)f(\v_2,\x_1+2a\kk,t).
\end{equation}
In the limit of vanishing filling factor the Enskog factor approaches unity,
 and we recover the \emph{Stosszahlansatz}.
 In the opposite
limit, when particles are so densely packed that random movement is
impossible, $Y$ diverges. This occurs when $\FF=0.74$ for a collection of identical
spheres arranged in a face-centred cubic lattice (AT86).

 The Enskog factor cannot be calculated from
within the compass of kinetic theory, but
 a number of expressions for it have been derived when the kinetic gas is elastic,
 non-shearing, and possessing a small filling factor (Ree and
Hoover, 1967; Carnahan and Starling, 1969; Devore, 1984). This paper
employs for $Y$
the convenient interpolation of AT86 to the
molecular dynamics data of Alder and Wainwright (Ree and Hoover,
1967). These details appear later.

\vskip0.3cm

We now present the collision frequency. 
Consider two particles with velocities $\v_1$ and $\v_2$ with the
first located at $\x$. For
a collision to occur characterised
 by a line of centres $\kk$
 in the solid angle $d\kk$,
 the second particle must be within a skew cylinder of volume $4 a^2
(\g\cdot \kk)d\kk dt$. Therefore the frequency of a collision
between two particles  with velocities in the volumes $d\v_1$ and
$d\v_2$, centred at $\v_1$ and $\v_2$,
with their line of centres
 within the solid angle $d\kk$ centred on $\kk$, and
 with particle 1 located in the small volume $d\x$
centred on $\x$,
is
\begin{equation} \label{rate} 
 4\,f^{(2)}(\v_1,\x,\v_2,\x+2a\kk,t) a^2 (\g\cdot \kk) d\kk\, d\v_1
 d\v_2 d\x.
\end{equation}
Into this expression we may substitute Eq.~\eqref{Ensk} when we are
ready.

\subsubsection{The collisional production integral}

The collisional rate of change per unit volume of some particle 
property $\psi$ is the integral (taken over all possible binary collisions)
of the change in $\psi$ in a particular collision, multiplied by the
probable frequency of such a collision. A factor of one half should
also be included, otherwise we will count every collision twice (once for
particle 1 and once for particle 2). 
We denote the resulting quantity by
$\textsf{C}(\psi)$ and express it as
\begin{equation} \label{colchange}
 \textsf{C}(\psi)= 2\int
H(\g\cdot\kk)\Delta(\psi)\, f^{(2)}(\v_1,\x,\v_2,\x+2a\kk,t) a^2 (\g\cdot \kk)
d\kk\,d \v_1 d \v_2,
\end{equation}
where the velocity integrations are over all space, the $\kk$
integrations are over the surface of the unit sphere, and $H$ is
the Heaviside step function (Jenkins and Richman, 1985). The step function ensures that the
integration only includes real collisions, i.e.\ ones that satisfy
$\g\cdot\kk>0$.
 Note importantly that
$\textsf{C}(\psi)$ is nonlocal in space, because production of $\psi$
at $\x$ necessarily depends on the values that certain fields take at
small but finite distances away.
To determine the collisional production of second moment, $\psi=\w\w$,
 we use the
expression~\eqref{2ndm} for $\Delta(\w\w)$. We subsequently define
$\mathbf{q}\equiv \textsf{C}(\w\w)$ (cf. Eq.~\eqref{qqq}).

If the system is dilute we set
$\FF\ll 1$ which forces $Y\approx 1$. Also we do not distinguish between
the centres of the two particles, i.e.\ we assume the mean flow and the distribution
function do not vary significantly on length scales of order $a$. Thus
$\hh\approx 0$ and $f(\v,\x+2a\kk,t)\approx
f(\v,\x,t)$. If we set $\psi=\w\w$, these simplifications reduce expression \eqref{colchange} to
that of Eq.~(24) in Goldreich and Tremaine (1978).

\subsubsection{The collisional flux of momentum}

Consider the momentum transferred in a collision from particle 1 to particle 2 
when they both straddle a surface of area $dS$ with normal
$\mathbf{n}$. The normal component of
momentum transferred across the surface from particle 1 to particle 2 
is then $m\,\jj\cdot\mathbf{n}$, and
the condition that the particles straddle the surface in the right
sense is
$\kk\cdot\mathbf{n}>0$.

Consider all the momentum transferred across the surface by such
 collisions. Particle 2 must lie just above the surface and sit in a
 skew cylinder of volume $a(\kk\cdot\mathbf{n})dS$. But the particle
 striking it must lie in the cylinder $4 a^2 (\g\cdot\kk)d\kk dt$ (for
 collisions within solid angle $d\kk$ and during time
 $dt$). Therefore, the average collision rate for particles
 with velocity ranges $d\v_1$ and $d\v_2$, straddling the surface
 $dS$, oriented in the range $d\kk$ with particle 1 at $\x$ is
\begin{equation}
  4 f^{(2)}(\v_1,\x,\v_2,\x+2a\kk,t) a^3 (\kk\cdot\mathbf{n}) (\g\cdot \kk) d\kk\, d\v_1
 d\v_2 dS,
\end{equation}
provided $\g\cdot\kk>0$ and $\kk\cdot\mathbf{n}>0$.
The total momentum flux density resulting from collisions we denote by $\PP$ and
define through
$$ \mathbf{n}\cdot\PP \equiv 4 m\,a^3 \int H(\g\cdot\kk) f^{(2)}(\v_1,\x,\v_2,\x+2a\kk,t) 
(\kk\cdot\mathbf{n}) (\g\cdot \kk)\,\jj\,d\kk \,d\v_1\,d\v_2,$$
thus
\begin{equation} \label{colmom}
\PP = 2 m\,a^3 \int H(\g\cdot\kk) f^{(2)}(\v_1,\x,\v_2,\x+2a\kk,t) 
(\g\cdot \kk)^2(1+\varepsilon)\kk\kk\,d\kk \,d\v_1\,d\v_2.
\end{equation}
If we assume that the mean flow $\u$ varies slowly across a particle diameter, then we may
write 
$$\hh= \u(\x,t)-\u(\x+2a\kk,t)\approx -2a\,\kk\cdot\nabla\u,$$ in which case
 we have the integral equivalent of Eq.~\eqref{enervate}:
\begin{equation}
\frac{1}{2}m\,\text{tr}(\mathbf{q})= -D - \PP:\nabla\u
\end{equation}
where the rate of dissipation is
\begin{equation}
D=\frac{1}{2}a^2\,m\int  H(\g\cdot\kk)
f^{(2)}(\v_1,\x,\v_2,\x+2a\kk,t)(1-\varepsilon^2)(\g\cdot\kk)^3 d\kk
d\v_1 d\v_2.\end{equation}
This relates the collisional production of energy to the collisional
momentum flux, and shows that the formalism is consistent; the
quantity $\PP:\nabla
\u$ is just the rate at which the collisional momentum flux extracts
energy from the mean flow.

\subsubsection{The peculiar velocity distribution function}

To make any progress in analysing the governing equations the integral
expressions must be simplified. To that end we propose a
specific model for the distribution function. First we introduce
the
peculiar velocity distribution function $f_{\w}(\w,\x,t)$ defined so that
$$
 f(\v,\x,t)=n(\x,t)f_{\w}(\v-\u(\x,t),\x,t).
$$
Following Goldreich and Tremaine (1978) and AT86,
a triaxial Gaussian model for $f_\w(\w,\x,t)$ is adopted:
\begin{equation}
f_{\w}= [ (2\pi)^3\,\text{det}(\WW)]^{-1/2}\text{exp}\left[
  -\tfrac{1}{2}\WW^{-1}:\w\w \right],
\end{equation}
where the dependence of $f_{\w}$ and $\WW$ on $\x$ and $t$ is
 understood. This form satisfies the required moment properties of
 Eqs \eqref{ch2.1}-\eqref{ch2.3}.

 To facilitate the manipulations of the next section we now
assume $\WW$ does not vary appreciably on lengthscales of order $a$.
This is acceptable if we only wish to analyse variations on long radial
 scales. In fact, if $a\sim H$ 
then the assumption is consistent with the closure scheme
in Section 2.1. Note that we have already assumed vertical isothermality, and
so $\WW$'s variation with $z$ is neglected.

\subsubsection{Transformation of the integrals}

First we transform from the variables $(\v_1,\v_2)$ to the relative
and centre of mass velocities, $(\g,\v_c)$, the latter defined by
$$ \v_c=\frac{1}{2}(\w_1+\w_2).$$
The Jacobian determinant of the transformation is unity and the only
appearance of $\v_c$ in the integrands is in 
$$ f_{\w}(\w_1)\,f_{\w}(\w_2)= \left[(2\pi)^3
  \text{det}(\WW)\right]^{-1}\text{exp}\left[
  -\WW^{-1}:\left(\v_c\v_c+\tfrac{1}{4}(\g-\hh)(\g-\hh)\right)\right].
  $$
This allows us to integrate over $\v_c$:
$$ \int \text{exp}\left[ -\WW^{-1}:\v_c\v_c\right] \,d\v_c=
 \left[\pi^3\text{det}(\WW)\right]^{1/2}.$$
What remains is
\begin{align}
\mathbf{q} &= 2a^2\int H(\g\cdot\kk)(\g\cdot\kk)\,
\widetilde{Y}(\x,\kk)\Delta(\w\w)f_{\g}(\g-\hh) \,d\kk\,d\g, \\
\PP &= 2a^3 m \int H(\g\cdot\kk) (\g\cdot\kk)^2\,
\widetilde{Y}(\x,\kk)\,(1+\varepsilon)f_{\g}(\g-\hh)\,\kk\kk
\,d\kk\,d\g,
\end{align}
where we have introduced, for notational brevity, the function
\begin{equation} \label{Ytilde}
 \widetilde{Y}(\x,\kk)=
Y\left[n(\x+a\kk)\right]\,n(\x)\,n(\x+2a\kk),
\end{equation}
and a distribution function for $\g$,
$$ f_{\g}(\g)= \left[(4\pi)^3\text{det}(\WW)\right]^{-1/2}\text{exp}
\left(-\tfrac{1}{4}\WW^{-1}:\g\g \right). $$
The latter is normalised such that $\int f_{\g}\,d\g=1$. Note that
$f_{\g}(\g-\hh)$ is an off-centred triaxial Gaussian, and the offset
$\hh$ which derives from the background mean flow, will depend on $\kk$.

We now tackle the $\g$ integrals. We express $\g$ space in Cartesian
coordinates $(X,Y,Z)$,
orienting them so $\eee_Z=\kk$. Thence we can remove the step function
and replace the integration domain with the half space, $\{(g_X,g_Y,g_Z):\,
g_X,\,g_Y \in \mathbb{R},\,\,
  g_Z\in\mathbb{R}^+\}$. This choice also simplifies the functional 
dependence of the coefficient of restitution to
$\varepsilon=\varepsilon(g_Z)$.
Consequently, the $g_X$ and $g_Y$ integrations are straightforward, and the $g_Z$
integrations go through if we introduce certain `averages' of
$\varepsilon$.
 The biggest problem is returning the result to
coordinate-free form, which requires some algebraic insight. 
But once this is done we have
\begin{align} 
\mathbf{q}&= 2\pi^{-1/2}a^2\int
\widetilde{Y}\,\zeta^{1/2}\times \notag\\
&\hskip1cm \left\{\, \zeta\langle\,(1+\varepsilon)^2\,\rangle_3
\,F_1\,\mathbf{K}-  [\,(1+\langle \varepsilon\rangle_3)\,F_1-
\xi(1+\langle \varepsilon\rangle_2)\,F_2\,]\, \mathbf{G}\, \right\} \,d\kk, \label{qdot} \\
\PP&=2\pi^{-1/2}a^3\int
\widetilde{Y}\, \zeta(1+\langle\varepsilon\rangle_2)\,F_2\,\mathbf{K}\,
d\kk, \label{pdot}
\end{align}
where the dyadics $\mathbf{K}$ and $\mathbf{G}$ are defined through
\begin{equation}
\mathbf{K}=\kk\kk, \qquad \mathbf{G}= (\WW\kk)\kk +\kk(\WW\kk),
\end{equation}
the $F$ functions are 
\begin{align} \label{F1}
F_1 & = 2(1+\xi^2)e^{-\xi^2}+\pi^{1/2}\xi\,(3+2\xi^2)(1+\text{erf}\,\xi), \\
F_2 & = \,2\xi\,e^{-\xi^2}+\pi^{1/2}(1+2\xi^2)(1+\text{erf}\,\xi),\label{F2}
\end{align}
where erf is the error function; the two variables $\zeta$ and $\xi$ are defined through
\begin{equation} \label{varsky}
 \zeta=\WW:\kk\kk, \qquad \xi=
\frac{1}{2}\frac{\hh\cdot\kk}{(\WW:\kk\kk)^{1/2}},
\end{equation}
and the average $\langle\psi\rangle_p$ by
\begin{equation} \label{fatavo}
\langle \psi \rangle_p(\zeta,\xi) = \frac{ \int_0^{\infty} x^p\,
  \psi(x\,\zeta^{1/2})\,\,e^{-x^2/4+\xi x} dx}{\int_0^{\infty} x^p
  e^{-x^2/4+\xi x}\,dx}, 
\end{equation}
for some general function $\psi(x)$. If $\psi$ is a constant then $\langle
\psi\rangle_p=\psi$. Expressions \eqref{qdot} and \eqref{pdot} are analogous to
Araki and Tremaine's (1986) but are more general as they include the
dependence of $\varepsilon$ on impact velocity. They are simpler,
requiring the evaluation of 
two-dimensional rather than four-dimensional integrals, and also benefit
from being coordinate free.

 The quantities that appear in Eq.~\eqref{varsky} have
 straightforward interpretations: $\zeta$ quantifies
 the magnitude of the squared impact velocity for a given orientation $\kk$,
 while $\xi$ measures the ratio of the shear velocity across a
 particle radius to the velocity
 dispersion, and is hence $\propto R$. This means dilute expressions
 may be recovered by letting $\xi=0$ and $\widetilde{Y}=n(\x)^2$.

In order to evaluate the remaining integrals, expressions for $\varepsilon$ and
$\hh$ must be supplied. The experimentally derived, piecewise power
law of \eqref{bridge} is the obvious choice for $\varepsilon$, while
we treat the velocity gradient as locally uniform, as before, i.e.
$ \hh= -2a\,\kk\cdot\nabla\u.$
Therefore:
$$\xi= -a\,\frac{(\eee:\kk\kk)}{(\WW:\kk\kk)^{1/2}} $$
where $\eee$ is the rate of strain tensor.

 In addition, we require the form of
$n(\x)$ and $Y[n(\x)]$. The Enskog factor may take one of the 
formulae mentioned, but the number
density should be determined from the vertical momentum
balance. Alternatively, an explicit form could be supplied. We discuss
these issues a little later.   

\vskip0.3cm

The integral terms as they stand, though still involved
 can be computed numerically without much difficulty.
 Nonetheless, additional
assumptions yield further useful simplifications.

First, we may suppose that
$n$ does not vary appreciably on scales of order $a$; so,
like $\WW$, it may be considered to be the same at the
position of each
particle in contact: i.e.\ $\x_1=\x_2=\x$. Then
\begin{equation} \label{Ytildeloc}
 \widetilde{Y}= Y[n(\x)]\,n(\x)^2
\end{equation}
and can be brought outside the integral.
This means that as far as the collision frequency is concerned
 nonlocality is neglected
and
 $n$ is constant; for
all other purposes (in the height equation, etc) $n$ varies and must
be determined from \eqref{verteqm}.
If we consider homogeneous equilibria or
 largescale horizontal variations, $n$ will only exhibit appreciable
variations in the $z$ direction (the disk thickness being of order
$a$) and so, essentially, it is this vertical variation that $\widetilde{Y}$ omits.
 We hence refer to the adoption of \eqref{Ytildeloc} as `vertical
locality', as opposed to the `vertical nonlocality' of the
full model (Eq.~\eqref{Ytilde}). Note that nonlocality in the shear,
introduced by $\mathbf{h}$, remains and is crucial.

 Second, we may suppose that for the purposes of
the collisional kinematics the coefficient of restitution is an
averaged quantity dependent on only macroscopic variables, principally
$c$, as was done in Latter and Ogilvie (2006). Hence, $\varepsilon$ is constant as far as
the integrations are concerned. 

Third, we may approximate the integrands by polynomial interpolation
 or series expansion. These approaches are valid if dense effects and/or
  anisotropy is small, and permit the remaining integrations to go through. We shall examine the last
  two assumptions in some detail.

\subsubsection{Averaging the coefficient of restitution}

As we have seen, the expressions for $\mathbf{q}$ and
$\mathbf{p}^{\text{c}}$ involve 
various averages of $\varepsilon$ which are described in Eq.\
\eqref{fatavo}. These can only be analytically evaluated for
relatively simple functions, and then the result is rather complicated. For
example, if we set the coefficient to be a straight power law, $\varepsilon=
(\g\cdot\kk/v_c)^{-p}$, we get
\begin{align} \label{firstav1}
\langle \varepsilon \rangle_2 &= \frac{2^{4-2p}}{\sqrt{\pi}}\,
\Gamma(2-\tfrac{1}{2}p)\Gamma(\tfrac{3}{2}-\tfrac{1}{2}p)\,\left(\frac{\sqrt{\zeta}}{v_c} \right)^{-p}
\cdot\left(\frac{e^{-\xi^{2}}}{F_2(\xi)}\right)\,H_{p-3}(-\xi), \\
\langle \varepsilon \rangle_3 &= \frac{2^{5-2p}}{\sqrt{\pi}}\,
\Gamma(\tfrac{5}{2}-\tfrac{1}{2}p)\Gamma(2-\tfrac{1}{2}p)\,\left(\frac{\sqrt{\zeta}}{v_c} \right)^{-p}
\cdot\left(\frac{e^{-\xi^{2}}}{F_1(\xi)}\right)\,H_{p-4}(-\xi), \label{firstav2}
\end{align}
where $H_{\nu}$ is the Hermite function, $\Gamma$ is
the gamma function (Abramowitz and
Stegun, 1965), and $F_1$ and $F_2$ are given by Eqs \eqref{F1} and \eqref{F2}.
 Unfortunately, more complicated functions, including the
more accurate piecewise law,
do not yield a closed form for $\langle\varepsilon\rangle_q$,
in which case the collision
terms require the numerical evaluation of three integrals each. 

We aim to employ the piecewise $\varepsilon$ law and,
 given that the collision terms are sufficiently complicated, it
would be convenient if we could eliminate the
third integral. Moreover, expressions like \eqref{firstav1} and
 \eqref{firstav2} are complicated functions of the integration variable
 $\mathbf{k}$ and hence computationally more intensive. 
These consideration motivate the
technique of `preaveraging'. Because
 the elasticity laws we possess are in some sense `the average' of a number of
experimental collisions anyway, philosophically, at least, this
approximation is justified.
In this subsection we provide details of a possible
preaveraging process. 

Let us define
$A_\gamma(\psi)$ as the weighted average of the function $\psi(\g\cdot\kk)$ over all
collisions. The weighting function is $(\g\cdot\kk)^{\gamma-1}$, and
$\gamma$ is a real positive number. We define
\begin{align} \label{avo}
A_\gamma(\psi)\equiv  \frac{ \int H(\g\cdot\kk)\,\, \psi(\g\cdot\kk)\,\,Y\, f_1 f_2\,\,
  (\g\cdot\kk)^\gamma\, d^2\kk d^3\w_1 d^3 \w_2}{\int  H(\g\cdot\kk)\,\,Y\, f_1 f_2\,\,
  (\g\cdot\kk)^\gamma\, d^2\kk d^3\w_1 d^3\w_2}
\end{align}
where $f_1$ and $f_2$ represent the peculiar velocity distribution function
evaluated at particle 1 and 2 respectively. Obviously, $A_\gamma(\psi)=\psi$
if $\psi$ is a constant. We now replace each instance of $\langle
\varepsilon \rangle_q$ with $A_q(\varepsilon)$, i.e.~
$\langle\varepsilon \rangle_q\approx A_q(\varepsilon)$. For a simpler but
less accurate formalism, we replace every instance of $\langle \varepsilon
\rangle_q$ with $A_\gamma(\varepsilon)$ for a \emph{fixed} $\gamma$
for all $q$.

 The integral \eqref{avo} can be
evaluated by the same techniques as earlier if it is
assumed that $f$ is a triaxial Gaussian. However, the result is relatively
complicated and it so happens there is little gain over
the fully consistent model of \eqref{qdot} and \eqref{pdot}.  
An acceptable
approximation is the adoption in \eqref{avo} of a Maxwellian, $f\propto
\text{exp}\left[-w^2/(2c^2)\right]$. However to complete all the
integrals we must adopt \eqref{Ytildeloc} and neglect the
nonlocal contributions, i.e.\ set $\xi=0$. This is less justifiable but mathematically
advantageous and, in fact, yields results in good agreement with the
full model, as we explore in Section 3.10. 

Implementing these approximations for
 the piecewise power
law function, $\varepsilon=\text{min}\{1,(\g\cdot\kk/v_c)^{-p}\}$, yields
\begin{align} \label{whoah!}
A_2(\varepsilon) &= \text{erf}\,X - \frac{2}{\sqrt{\pi}}\,X\,\left(1+\frac{1}{\mu_2}\,X^2\right)\,e^{-X^2}
- \frac{2}{\mu_2\sqrt{\pi}}\,X^5\,E_{\mu_2}(X^2)  \\
A_3(\varepsilon) &= 1- \left(1+X^2-\frac{2}{\mu_3}\,X^4 \right) e^{-X^2} -\frac{1}{2\mu_3}\,X^6\,E_{\mu_3}(X^2) \label{whoah!2}
\end{align}
where $X= v_c/(2c)$, $\mu_s=(p-1-s)/2$, and $E_\mu(x)$ is the exponential integral
E function. Though they may seem a little complicated, these expressions are especially helpful because they
 involve only the squared velocity
dispersion ($c^2$) and none of the integration variables (through $\zeta$ and
$\xi$). Consequently, they can be brought out of the integrals entirely.

\subsubsection{Interpolating the integrands}

Because the $F$ functions rapidly increase
with $\xi$, their Taylor series (truncated at a manageable order) 
are poor approximations unless $\xi$ is very small. But dense ring equilibria fall 
into the regime of $|\xi|\sim 1$. An alternative is to interpolate the
$f$'s on some range of $\xi$, which we can estimate from
Eq.~\eqref{varsky}. 

To proceed, vertical locality is assumed, so that
$\widetilde{Y}=Y(\x)\,n^2(\x)$, and $\varepsilon$ is treated as
preaveraged, with the additional approximation,
$A_2(\varepsilon)=A_3(\varepsilon)$ in the collisional
production integral. 
Enforcing these gives
\begin{align*}
 \mathbf{q} &= 2\pi^{-1/2}\, a^2\, \widetilde{Y}[1+A_3(\varepsilon)]\left\{
[1+A_3(\varepsilon)]\int\zeta^{3/2}F_1\mathbf{K}\,d\kk-\int\zeta^{1/2}F_3\mathbf{G}\,d\kk\right\},\\
\PP &= 2\pi^{-1/2}a^3
\widetilde{Y}[1+A_2(\varepsilon)]\,\int \zeta\,F_2\,\mathbf{K}\,
d\kk,
\end{align*}
where $F_3\equiv F_1-\xi F_2 $.

We now interpolate the $F$ functions. For small anisotropy we have the upper bound
$$ |\xi|< \frac{3}{2}\,R\,\left\|\frac{\mathbf{e}}{\Omega}\right\|,$$
which for equilibrium solutions with $R<1$ gives $|\xi|<9/8$.
On this range we find that $F_1$ can be successfully 
interpolated by a cubic in $\xi$, and $F_2$ and $F_3$ by quadratics:
\begin{align} \label{interpen}
F_1= \sum_{n=0}^3 F_{1n}\, \xi^n, \qquad F_2= \sum_{n=0}^2 F_{2n}\, \xi^n,
\qquad
 F_3 = \sum_{n=0}^2 F_{3n}\, \xi^n.
\end{align}
  where the $F_{jn}$ are pure numbers arising from
 the interpolation range and the interpolation method.

Next, we express $$ \WW= c^2(\mathbf{1}+\mathbf{\ani})$$
where the squared velocity dispersion is $c^2=\text{Tr}(\WW)/3$ and
$\ani$ is the anisotropy tensor, which is symmetric and
traceless. It follows that
$$\pib= -mn\,c^2\,\ani$$
if the viscous stress is denoted by $\pib$.
Because $\|\ani\|<1$, the integrands may be systematically expanded
 in $\ani$ through
$$\xi=-\frac{a}{c}\,(\eee:\kk\kk)\left[\,1-\tfrac{1}{2}\ani:\kk\kk+\tfrac{3}{4}(\ani:\kk\kk)^2+\mathcal{O}(\ani^3)\right],$$
and
$$ \zeta = c^2\,\left[ 1+ \mathbf{a}:\kk\kk\right].$$
 
Now we are in a position to evaluate the integrals directly.
 Expressions for $\mathbf{q}$ and $\mathbf{p}^c$ may be written in terms of
 spherical surface
integrals of the form
$$ J_{i_1 i_2\dots i_{2l}}= \frac{1}{4\pi}\int_S k_{i_1} k_{i_2} \dots
k_{i_{2l}}\, d\kk,$$
where $S$ is (the surface of) the unit sphere. Necessarily
$$ J_{i_1 i_2\dots i_{2l}}= \frac{1}{2l+1}\, I_{i_1 i_2\dots i_{2l}},$$
where 
$$ I_{i_1 i_2\dots i_{2l}}= \delta_{(i_1 i_2}\delta_{i_3 i_4} \dots
\delta_{i_{2l-1} i_{2l})},$$
and is the unique, totally symmetric, isotropic tensor of rank
$2l$. Parantheses around a set of $j$ indices denote the sum of all
permutations of the indices divided by $j!$ 

 The closed forms
of the integrands to $\mathcal{O}(\ani)$ are presented in Appendix A,
and, though complicated, are algebraic functions. As such, they possess a
great numerical advantage over previous kinetic approaches, and
certainly pave the way for dynamical analyses, particularly nonlinear
studies, which have been the sole preserve of hydrodynamics.

The main errors that characterise this formalism arise from
truncation in $\ani$. Thus, significantly anisotropic systems may not
be well described (for instance, rings suffering very few
collisions). In addition, when the velocity dispersion is extremely low or the rate
of strain large, the interpolation range must be extended, and at
some point
the interpolating polynomials may be of too low an order to adequately
approximate $F_1$, $F_2$, and $F_3$. This should only be an issue for some nonlinear
calculations, however.

\subsubsection{Expanding the integrands}

We have argued that a real ring lies in the regime $\xi\sim 1$, 
but it is nevertheless interesting to examine `warm' and dilute rings for
which $\xi$ is small. Doing so allows us to connect our formalism to other
well-known kinetic models, such as those of Goldreich and Tremaine (1978), and
Jenkins and Richman (1985).
 
 We expand the integrands of $\mathbf{q}$ and $\PP$ first in powers of $\xi$
and then in $\ani$, assuming both are small. All the integrals then can
be performed much as before. In the interest of space we omit these details. 

The closed algebraic expressions for $\mathbf{q}$ and $\PP$ so obtained
 are analogous to, and possibly
more systematic than, those computed by Jenkins and Richman (1985), who employ
Grad's prescription for the single particle distribution function
 (Grad, 1949).
 They concentrate on the case when
    the spatial derivatives of the mean fields are small, and the
    distribution nearly isotropic. In the language of our model,
    they examine orders up to and including $(\xi^0,\ani^1)$ and
    $(\xi^1,\ani^0)$, and then in the hydrodynamic limit, $\omega_c$ large.
    This permits them to obtain constitutive
    expressions for the hydrodynamic transport coefficients --- shear
    and bulk viscosities and thermal conductivity. Interestingly we
    can recover precisely the same expressions, despite the difference in
    our choice of distribution function. 

The leading terms, of order $\xi^0$, are just the collision integrals for
 a dilute gas. And because of the triaxial Gaussian ansatz adopted they
 coincide with the
collision term of the Goldreich and Tremaine (1978) model up to
whichever order in $\ani$ we choose to truncate. After
 vertical averaging we obtain
$$ \mathbf{Q}= \frac{8}{\pi}\,\tau\Omega N \, c^2
(1+\varepsilon)\left[-\tfrac{1}{3}(1-\varepsilon)\unit
-\tfrac{1}{5}(3-\varepsilon)\ani+\mathcal{O}(\ani^2)\right].$$
In contrast, the model of Shu and Stewart (1985) gives
$$ \mathbf{Q}=\frac{8}{\pi}\,\tau\Omega N\,c^2
(1+\varepsilon)\left[-\tfrac{1}{3}(1-\varepsilon)\unit
-\left(\frac{1}{1+\varepsilon}\right)\ani\right],$$
and their interpretation of the `dilute' H\"{a}meen-Anttila model
(H\"{a}meen-Anttila, 1978, 1981) is
$$ \mathbf{Q}=\frac{8}{\pi}\,\tau\Omega N\,c^2
(1+\varepsilon)\left[-\tfrac{1}{3}(1-\varepsilon)\unit
-\tfrac{1}{6}(3-\varepsilon)\ani\right].$$
Here upper case indicates vertical averaging, as before.
All these models agree at $\mathcal{O}(\ani^0)$ but disagree
at $\mathcal{O}(\ani)$. However, the difference at this order between
the Goldreich and Tremaine and H\"{a}meen-Anttila models is only a small difference
in the constant coefficient ($1/6$ against $1/5$).
This accounts for the (relatively) good agreement these two models enjoy in equilibrium
calculations (Stewart \emph{et al.}, 1984).

  \section{Equilibria}

Now that the various models have been sketched, we employ them to calculate
 dense ring equilibria.
  We solve for the vertically
 averaged pressure tensor and the disk height
  for a given vertical structure of $n$. 
First, we compare the results predicted by our kinetic model with
those produced by the kinetics of AT86 and the $N$-body
 simulations of Wisdom and Tremaine (1988) and Salo (1991).
Second, we test the
appropriateness of the three models introduced 
 in the preceding sections, namely,
 that of `vertical locality', the `preaveraged'
 $\varepsilon$ model, and the interpolation model. Because these
 formalisms significantly simplify the problem it is worthwhile
 testing the approximations upon which they are founded.
 
\subsection{Governing equations}

We calculate the steady homogeneous state of a disk rotating about a
perfectly spherical planet and whose self-gravity is negligible in the
horizontal balances. In
the shearing sheet this equates to
$n=n_0(z)$ and $\u=-(3/2)\Omega\,x\,\mathbf{e}_y$. 
The equilibrium pressure tensor equations are
\begin{align}\label{eqm1}
4\Omega W_{xy} + Q_{xx}/N&=0, \\
-\frac{1}{2}\,\Omega W_{xx}+2\Omega W_{yy}+ Q_{xy}/N  &=0, \\
-\Omega\, W_{xy} +  Q_{yy}/N &=0, \\
Q_{zz} &=0,\label{eqm2}
\end{align}
where, as earlier,  upper case indicates vertical averaging (except in the case of
$W_{ij}$ and $c$ which we assume independent of $z$). The natural
nondimensionalisation of this system is founded on particle radius
and orbital frequency, so that
$$t= \Omega^{-1}t^*, \quad x= ax^*,\quad \u=a\Omega \u^*, \quad
W_{ij}= a^2\Omega^2 W_{ij}^*,$$ 
where asterisked quantities are dimensionless.

In addition to this set of equations we need expressions for $n_0(z)$, $Y[n]$ and
$\varepsilon$. The last two can be supplied by simulations or experiments. The
first can be determined from the vertical hydrostatic balance
\eqref{verteqm}. We visit each in turn.

\subsection{The Enskog factor}

The Enskog factor is modelled according to Araki and Tremaine's
 prescription:
 a polynomial fit to the molecular dynamics data of Alder and
Wainwright (in Ree and Hoover, 1967)\footnote[2]{It
  should be noted that the values given by Araki \& Tremaine (1986)
  contain a typographical error. Their coefficient $D_3$ must read 0.8874 not 0.08874
 (Juergen Schmidt, private communication).}:
\begin{equation} \label{FF}
Y(\FF)=\sum_{i=0}^5\, D_i \FF^i
\end{equation}
where
\begin{align*}
D_0= 1.0000, \qquad  D_1=3.5496, \qquad D_2=-18.816 \\
D_3= 165.30, \qquad D_4= -407.27, \qquad D_5= 406.94
\end{align*}
Though strictly applicable to only non-shearing gases of elastic spheres, this
prescription should capture the qualitative behaviour of the Enskog
factor in an inelastic system (see AT86).

\subsection{The coefficient of restitution}

 For $\varepsilon$ we mainly use the piecewise power law of \eqref{bridge}
which has been experimentally derived by Bridges \emph{et al.} (1984),
Hatzes \emph{et al.}\ (1988), and Supulver \emph{et al.}\ (1995) for
sufficiently frosted ice particles in Saturnian conditions.
 The two parameters $p$ and $v_c$ we let vary, though recognising that 
Bridges' data give their values as $0.234$ and $0.0077$ cm s$^{-1}$,
Hatzes' data $0.20$ and $0.025$ cm s$^{-1}$, and Supulver's $0.19$ and $0.029$
cm s$^{-1}$. The preaveraged $\varepsilon$'s take the functional form of
\eqref{whoah!} and \eqref{whoah!2} in the collision terms. 
For one calculation we employ the power law form of $\varepsilon$, for which
we use the fully consistent expressions \eqref{firstav1} and
\eqref{firstav2}. 
These forms introduce the important dimensionles quantity $\textsf{R}_c\equiv
a\Omega/v_c$ which controls to some extent the equilibrium
velocity dispersion of the particles.

\subsection{The height equation and vertical locality}

 Instead of solving for $n$ a convenient alternative is to assume it takes a specific
 form (AT86), either 
 a Gaussian or polytrope. In
this case Eq.~\eqref{verteqm} becomes an algebraic equation for $H$.
This is then solved with the set \eqref{eqm1}--\eqref{eqm2}.

In the case of Gaussian vertical structure, we write, 
\begin{equation} \label{NNN}
n_0(z)= \widetilde{n}\,\,\text{exp}\left[-\frac{z^2}{2H^2}\right],
\end{equation}
where $H$ is the disk scale height and
$\widetilde{n}$ a constant. Consequently we can determine the optical depth from
\begin{equation} \label{tauie}
\tau = \frac{3\sqrt{2\pi}}{4}\,\,\frac{H}{a}\,\FF_0,
\end{equation}
where $\FF_0= \FF[\,\widetilde{n}\,]$, i.e.\ the filling factor in the
midplane ($z=0$). This expression gives flesh to the estimate
\eqref{nlest3}.

After multiplication by $z$ and vertical integration, the hydrostatic balance
 \eqref{verteqm}
 becomes
\begin{equation} \label{HHH}
\Omega^2\,H^2\left[1+
  \tfrac{3}{2}\,\sqrt{2}\,G_s\,\FF_0\right]-W_{zz}-P^c_{zz}/N=0,
\end{equation}
 where $N=\sqrt{2\pi}\,H\widetilde{n}$ is
the surface number density and
$$ G_s\equiv \frac{G\,m}{\Omega^2\,a^3}= \left(\frac{\rho_p}{\rho_S}\right)\left(\frac{r}{r_S}\right)^3,$$
expresses the disk self-gravity. Here $\rho_p$ and $\rho_S$ are the internal 
mass densities of a
particle and the planet respectively, $r$ is distance from the centre of
the planet, and $r_S$ is the radius of the planet itself.
 In Saturn's case, $\rho_S= 618\,\text{kg}\,\text{m}^{-3}$ and $r_S=6.03\times
 10^{4}\,\text{km}$; we then have
\begin{equation}\label{Gss}
G_s =
0.735\times\left(\frac{r}{10^{5}\,\,\text{km}}\right)^3\left(\frac{\rho_p}{100\,\,\text{kg}\,\text{m}^{-3}}\right).
\end{equation}
For particles composed of solid ice, $\rho_p=900\,\text{kg}\,\text{m}^{-3}$,
the parameter $G_s$ varies from about 6 at the inner edge of the B-ring, to 7.7
 at the outer edge of the B-ring, to approximately 9 at the outer edge of the
 A-ring. Of course, these estimates are rather rough. 
The internal density of real ring particles may depart signficantly
 from that of solid ice: particles may, for instance, be far more `fluffy'
 and
 possess a density less than that quoted (see, for example, Weidenschilling
 \emph{et al.}, 1984).

A polytropic model can be described by
\begin{equation}\label{polypoly}
 n_0(z)= \widetilde{n} \left( 1- \frac{z^2}{H^2} \right)^\alpha 
\end{equation}
for a given constant $\alpha$. Then the optical depth is
\begin{equation} \label{HHH2}
\tau = \frac{3\sqrt{\pi}\,\,\Gamma(1+\alpha)}{4\,\,\Gamma(3/2+\alpha)}\,\,\frac{H}{a}\,\FF_0
\end{equation}
and the vertical hydrostatic balance differs little in form from \eqref{HHH} except for the
addition of the new parameter $\alpha$,
\begin{equation} \label{verteqm2}
\,\Omega^2_0\, H^2\left(\tfrac{1}{2} (\alpha+\tfrac{3}{2})^{-1}  +\tfrac{3}{2}\, 
  g(\alpha)\,G_s\,\FF_0 \right) -W_{zz}- P^c_{zz}/N = 0,
\end{equation}
where
$$ g(\alpha)=\frac{2^{2\alpha+1}}{\sqrt{\pi}(\alpha+1)}\cdot
\frac{\Gamma(\alpha+\tfrac{3}{2})^2}{\Gamma(2\alpha+\tfrac{5}{2})}\,.$$
 This approximation has been used
by Borderies, Goldreich and Tremaine (1985) and Mosquiera (1996) (with
$\alpha=1$) and
is more accurate than the Gaussian model at describing monodisperse disks at moderate and large filling factors, as
we shall see.
 A problem, however, is the specification of $\alpha$ which
may vary with $\FF_0$ but which cannot be determined from the
governing equations. That said, our equilibrium solutions show that the equilibrium
quantities are relatively insensitive to the precise choice of $\alpha$.

\vskip0.3cm

If we assume vertical locality, the scale height $H$ can be removed from the
viscous stress equations \eqref{eqm1}--\eqref{eqm2} completely. Moreover we can
solve for $H$ explictly in \eqref{HHH} or \eqref{verteqm2} once the viscous stress
components are known, because then $\overline{\widetilde{Y}}/N$, and thus
$P^\text{c}_{zz}/N$,
 is independent of
$H$. In the case of the vertical Gaussian, the averaged $\widetilde{Y}$ is
\begin{equation} \label{verty}
\overline{\widetilde{Y}}/N= \frac{3}{16\pi\,a^3}\left(\sum_{j=0}^5
  D_j\,\FF_0^{j+1}/\sqrt{j+2}\right).
\end{equation}
In the case of the polytrope, we compute
\begin{equation}\label{verty2}
\overline{\widetilde{Y}}/N= \frac{3}{16\pi\,a^3}\left(\sum_{j=0}^5
  D_j\,\FF_0^{j+1}\, \frac{\Gamma(\alpha+3/2)\cdot\Gamma(j\alpha+1)}{\Gamma(\alpha+1)\cdot\Gamma(j\alpha+3/2)}\right).
\end{equation}

If we do not adopt vertical locality then $H$ appears in all the vertically averaged 
equilibrium equations through the vertically averaged $\widetilde{Y}$ (cf. Eqs \eqref{qdot}
and \eqref{pdot}). The height equation must then be solved concurrently with
\eqref{eqm1}--\eqref{eqm2}.
  Unfortunately, the vertical integration can only be
accomplished analytically for the Gaussian model. 
Usage of the polytrope model
must be accompanied by the assumption of vertical locality, unless we compute
the extra integral numerically. 

In the
Gaussian case, vertical nonlocality engenders the additional factor
$\text{exp}\left[-(a^2/H^2)k_z^2\right]$ in \eqref{verty}. This `weighting function'
 diminishes the effect
of those
collisions
in which the particles inhabit different vertical strata, or, put another way,
are oriented `more vertically'  with
respect to each other (exhibiting larger $k_z$). It follows that the factor
takes values between 1 (`perfectly horizontal' collisions) and
$\text{exp}[-a^2/H^2]$ (`almost vertical' collisions).
When the disk is thick (meaning, $a/H$ is small) then this range is very
narrow and the factor is essentially 1; vertical nonlocality is then unimportant.
 In the extremely thin limit,
i.e. $a/H\sim 1$, the factor can be smaller than $e^{-1}$, which expresses the
simple fact that in a near monolayer it would be unusual to find a pair of particles
inhabiting very different vertical positions. 

Finally we present an expression for the collision frequency. This should
scale roughly like $\pi a^2\,c\,Y n$. After inspecting \eqref{rate}, we settle on 
\begin{equation} \label{omegac}
\omega_c= 16\pi^2\,a^2\,c \left(\overline{\widetilde{Y}}/N\right),
\end{equation}
in which the constant coefficient is chosen so that the expression conforms with
the dilute $\omega_c$ calculated by Shu and Stewart (1985). In this limit the
vertical structure is Gaussian and we obtain a simple linear dependence on
optical depth: $\omega_c= (8/\pi)\Omega\,\tau$.

\subsection{Numerical technique}

 The equilibrium solution requires
 the roots of the algebraic system
 \eqref{eqm1}--\eqref{eqm2}; if vertical nonlocality is included we must add
 the
 the height equation \eqref{HHH} or
 \eqref{verteqm2} to this set.
 As these equations involve
 integral functions of $W_{ij}$ a number of numerical techniques are needed.
 The solutions
are obtained by using the multi-dimensional Newton's method, while the
 spherical surface integrations are
broken into azimuthal angle, which are accomplished using the trapezoidal
rule, and meridional angle, which are accomplished using Gaussian
 quadrature. \emph{Mathematica} was the programming language used, and
 the special functions that appear in the integrals
 are evaluated using the \emph{Mathematica} routines.
If $\varepsilon$ is specified, 
as either a constant or a function of $\g\cdot\kk$, the equilibrium solution
$W_{ij}$ (and $H$)
depends only on the parameter $\FF_0$.

\subsection{Comparison with the model of Araki and Tremaine (1986)}

In order to check our model
 we compared its predictions with
those of AT86, who adopt a constant $\varepsilon$,
 Gaussian vertical structure and vertical
nonlocality. If we do the same,
 the two models should
produce identical results because they share exactly the same assumptions.

We computed equilibrium properties
 with our formalism and checked it against those produced by the Araki and Tremaine
model (Schmidt, private communication). The parameters used were
 $\varepsilon=0$ and $G_s=7.4$, on one hand, and $\varepsilon=0.5$ and $G_s=0$
 on the other. Agreement to four significant figures was obtained and this
 seemed only limited to the accuracy chosen for the numerical integrations.
These results prove that the formalism we develop in this paper is
 equivalent to that of AT86.

\begin{figure}[!ht]
\begin{center}
\scalebox{.75}{\includegraphics{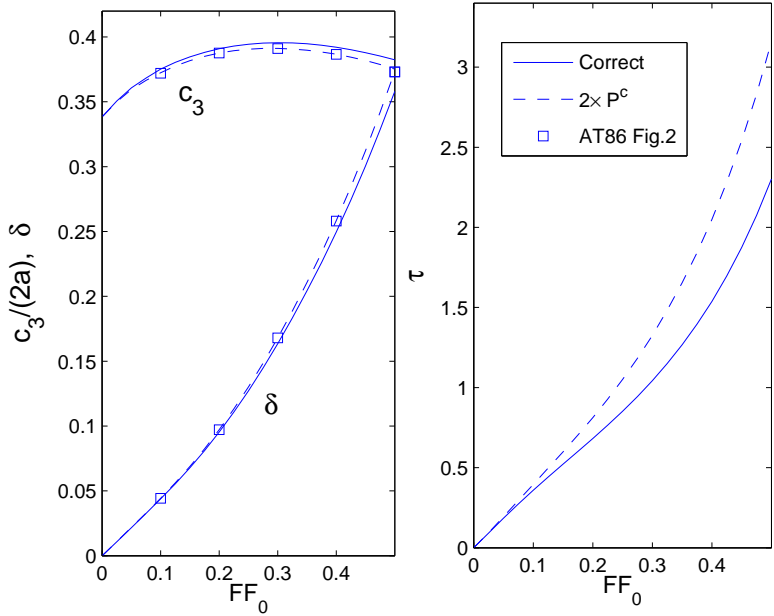}}
\caption{In these figures we show the effect on certain equilibrium properties
when a factor 2 is introduced into the collisional transport term, $P_{zz}^c$.
Additionally, we plot points from Fig.~2 in AT86
 in order to demonstrate that their computational routine suffers the
factor 2 error. In subplot (a), we present the orientation angle of the
velocity ellipsoid $\delta$ and the vertical velocity dispersion $c_3$. The
parameters taken are $\epsilon=0$ and
$G_s=7.4$. In subplot (b), we present optical depth $\tau$. The parameters are $\epsilon=0.5$ and $G_s=0$.
Generally the discrepancy is small, except in $\tau$ and $H$ when
self-gravity is weak.
}\end{center}
\end{figure}

It should be mentioned that the kinetic equilibria in Araki and Tremaine
(1986), and also Wisdom and Tremaine (1988), were computed by a routine which
introduces an erroneous factor of two in the calculation of $P_{zz}^c$ (given by their
Eq.~(142)).
 As a consequence their
computations overestimate the amount of collisional momentum transport in the
vertical direction. In the self-gravitating case $G_s=7.4$, which appears in
AT86, the discrepancy is small but noticeable. We
illustrate the discrepancy in Fig.~1a and also plot some points from
Araki and Tremaine's Fig.~2. As is clear, the latter sit perfectly on the
calculation that doubled $P_{zz}^c$.
In contrast, when self-gravity is omitted, the overestimated
collisional flux has greater influence, but only on those quantities which
depend closely on the vertical problem, namely optical depth $\tau$ and scale
height $H$. This is demonstrated in Fig.~1b where we plot $\tau$ versus
$\FF_0$ for the case $\varepsilon=0.5$ and $G_s=0$. For large $\FF_0$ the
discrepancy can be as large as $35\%$.

\subsection{Comparison with the simulations of Widsom and Tremaine (1988)}
   
The discrepancy just mentioned means that the agreement 
between the simulations of Wisdom and Tremaine (1988) and the AT86 kinetics
with respect to $\tau$ no longer holds, cf. their Fig.~15. 
We stress, though, that the good agreement witnessed by the
 `horizontal' equilibrium properties,
 (as functions of $\FF_0$) remain. 

The disagreement in $\tau$ we blame on the
assumption of the Gaussian vertical profile, which is appropriate at low $\FF_0$
but less so at higher filling factors, when the disk establishes a vertical
structure akin to a polytrope (see Fig's 3, 4, 8, and
9 in Wisdom and Tremaine, 1988). These two regimes and
the transition between them can only be captured by a
kinetic model that properly solves the vertical hydrostatic equilibrium, which
is precisely what is done in Araki (1991). The agreement with Wisdom and
Tremaine's simulations in this work is excellent, though the kinetic theory is
computationally expensive and suffers numerical difficulties. 

As an alternative to the approach of Araki (1991), we apply a model in
which polytropic vertical
structure is assumed from the outset. Such a model should
 approximate well the middle to larger filling factor regimes and tolerably
 approximate the low filling factor regime, thus saving us the arduous
 task of computing the vertical hydrostatic balance exactly.           
            
Subsequently, we computed equilibria using the polytropic formalism embodied in
  \eqref{polypoly}-\eqref{verteqm2}, and \eqref{verty2} using a fixed
 $\alpha$ and $\varepsilon=0.5$ and $G_s=0$. Much better
agreement was obtained with the Wisdom and Tremaine simulations at middling and larger filling
factors and reasonable agreement at lower filling factors (cf. Fig.~2).
Indeed, Fig.~2a nicely illustrates the transition between the two regimes ---
Gaussian at lower $\FF_0$ and polytropic at middle to large $\FF_0$.
 The transition occurs around $\FF_0=0.275$ or about $\tau=1$. Additionally, we
 plot the velocity dispersion versus $\FF_0$ to show, in contrast, that the horizontal
equilibrium properties are little altered by the choice of vertical model. 

\textbf{Fig.~2}
\begin{figure}
\begin{center}
\scalebox{.75}{\includegraphics{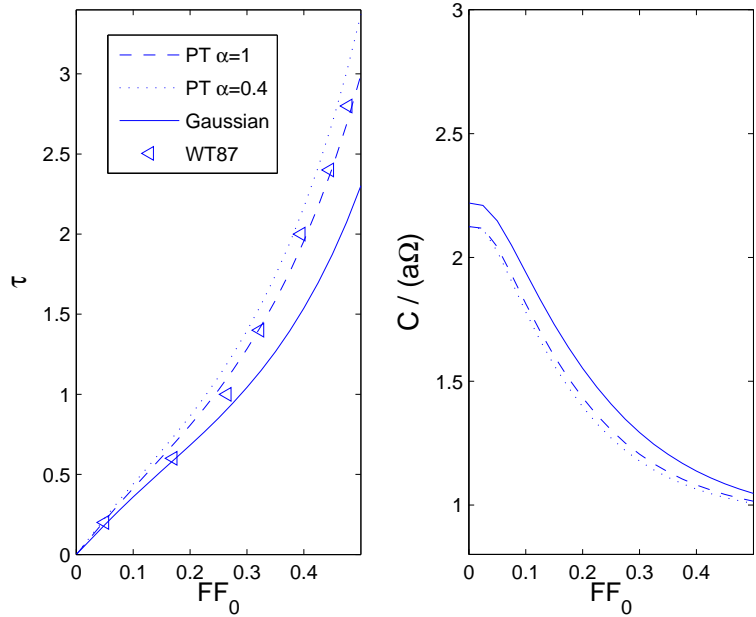}}
\caption{(a) Here we plot optical depth $\tau$, as a function of central filling factor $\FF_0$,
computed by various methods.
 The dashed and dotted lines correspond to
 kinetic models with polytropic vertical structure (`PT') and $\alpha=0.4$ and
 $\alpha=1$ respectively. The solid line corresponds to a kinetic model with
 Gaussian vertical structure. The diamond points (WT88) correspond to the simulations
 of Wisdom and Tremaine (1988), read from their Fig.~15. (b) The second panel
 presents 
 velocity dispersion $C$, as a function of
 central filling factor, computed by the three kinetic models. 
The coefficient of restitution is set to $0.5$ and $G_s=0$ in all cases.
}
\end{center}
\end{figure}

 The range of $\alpha$ that we employed was informed by simple inspection
of the vertical profiles of Wisdom and Tremaine (1988). These indicated that $\alpha$
takes values between $1$ to about 0.4 (and perhaps lower). It is clear,
though, that
changing $\alpha$ does not significantly alter the equilibrium properties, certainly not for
$W_{ij}$, as can
be seen in Fig.~2b. Thus it appears the polytropic model constrains the
equilibrium behaviour to a separate
regime effectively independent of the specific choice of $\alpha$: this means we can't
just produce any result we
like by tweaking $\alpha$. However, the best agreement is obtained with
$\alpha=0.9$ and we employ that value throughout the rest of the paper.

\subsection{Comparison with Salo (1991)}

We next compare our equilibrium results with those of Salo's (1991)
 simulations for which the coefficient of restitution varies as a function of impact
 velocity. This supplies us with results with which to test the kinetic
 model's treatment of nonconstant $\varepsilon$.
  Moreover, this set of data is particularly useful as it
 includes a number of runs with different $v_c$, one of the
 key parameters in our model. The disk, otherwise, 
is non-self-gravitating, and
 particles are assumed to possess a radius of $100$ cm and orbit at the
 location in the B-ring where $a\Omega= 0.02$ cm s$^{-1}$.
 Collisions dissipate energy according to the piecewise power law,
 Eq.~\eqref{bridge}, with $p=0.234$ and $v_c$ a multiple of $v_b=
 0.01$ cm s$^{-1}$. By varying $v_c$ the parameter $\textsf{R}_c$ can be
 controlled,
 which roughly governs
 the velocity dispersion and the `denseness' or `diluteness' of the gas,
 at least for intermediate collision frequencies. 

 We will actually compare our results with
 more recent equilibrium simulations which reproduce the runs of Salo (1991)
but with a five-fold increase in particle number and duration (Salo, private
 communication). The results are much the same but with reduced scatter.

The kinetic model adopted in this section employs
\begin{itemize}
\item a preaveraged $\varepsilon$, the piecewise power law, using \eqref{whoah!} and \eqref{whoah!2}
\item polytropic vertical structure with $\alpha=0.9$
\item vertical locality.
\end{itemize}
 When nondimensionalised
the model depends on $p$, $\textsf{R}_c= a\Omega/v_c$, $G_s$, and
 $\FF_0$. In this section $G_s=0$ and $p=0.234$ as mentioned, and $\FF_0$ and
 $\textsf{R}_c=2(v_b/v_c)$ will vary. Once supplied with these inputs
our algorithm produces the nondimensionalised
 $W_{ij}$ and then computes a number of related properties: the viscosities,
$$ \nu^L= \frac{W_{xy}}{3\Omega/2}, \qquad
\nu^{NL}=\frac{P^c_{xy}/N}{3\Omega/2},$$
 the disk semithickness $H$, optical depth $\tau$, and preaveraged
coefficient of restitution, $A_3(\varepsilon)$. Throughout this section the
quantity $A_3(\varepsilon)$ will be referred to as simply $\varepsilon$.

\subsubsection{Results}

In Fig.~3 we plot the equilibrium velocity dispersion, the disk
scale height, the averaged $\varepsilon$, and the central filling factor
against $\tau$ for various $v_c$. In Fig.~3a we also
 include the results of Salo's simulations.

 As is clear from Fig.~3a
 the kinetics and the simulations are consistent,
and match particularly well when filling factor is
 high. In general, the kinetic model slightly overestimates the velocity
 dispersion, particularly at low optical depths.
 The preaveraging assumption probably contributes most heavily to this
 discrepancy. Errors arising from the asumptions of vertical
 locality should be secondary because the disk thickness $H$ is relatively large
 (cf. Fig.~3c). But the choice of polytropic
 vertical structure may play some part, even though Fig.~2b suggests
 that a Gaussian model would also overestimate $C$. For
large ratios $v_c/v_b$ --- when the velocity dispersion is large and $\FF_0$ low
 (see Fig.~3d) --- the vertical disk structure will probably resemble a
 Gaussian not a
 polytrope.

These considerations aside, the general trend is quite clear: as $v_c$
 increases so does the equilibrium velocity dispersion. This
admits a relatively straightforward interpretation. Because collisions
 must dissipate energy to offset viscous heating, 
 $C$ must be at least $\mathcal{O}(v_c)$: if $C$ were much less than
 $v_c$ very little energy at all would be lost in collisions and
 certainly not enough to maintain an energy balance. It must follow that if
 $v_c$ is increased so must the equilibrium $C$.

 \begin{figure}[!ht]
\begin{center}
\scalebox{.38}{\includegraphics{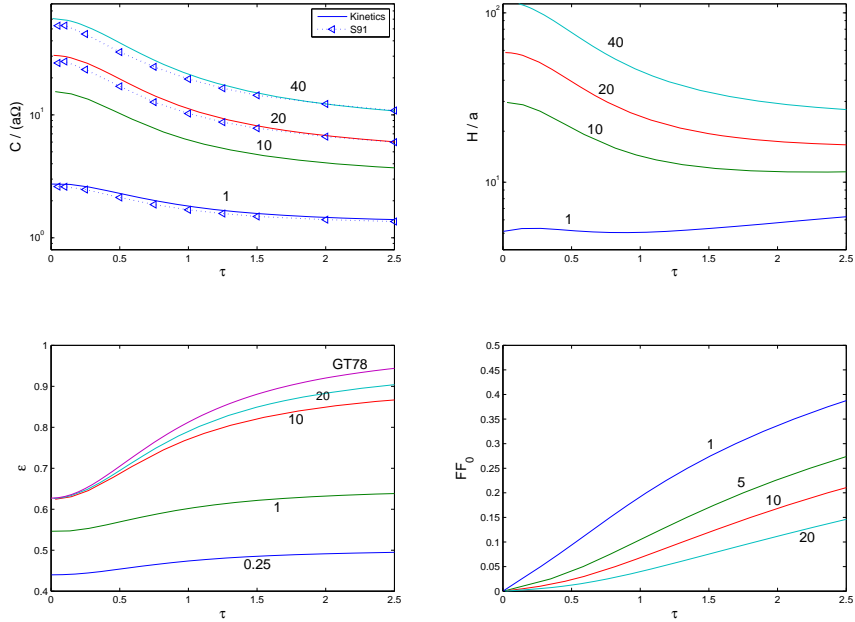}}
\caption{Various equilibrium properties computed by a polytropic kinetic model
($\alpha=0.9$) with a
preaveraged $\varepsilon$ (solid curves), and by the simulations of Salo
(1991) which we denote by `S91'
(triangles).
 We employ different ratios of $v_c/v_b$ (inserted
  adjacent to the relevant curve) but fixed $p=0.234$. 
 The quantities are plotted as
functions of optical depth $\tau$. The four panels correspond to:
 (a) equilibrium velocity dispersion $C$, (b) disk scale
  height $H$, (c) averaged equilibrium coefficient of restitution
  $\varepsilon$, and (d) central filling factor $\FF_0$.
   The Goldreich and Tremaine
  (1978) dilute disk energy balance is referred to as GT78 in subplot (c).}
\end{center}
\end{figure}

 \begin{figure}[!ht]
\begin{center}
\scalebox{.35}{\includegraphics{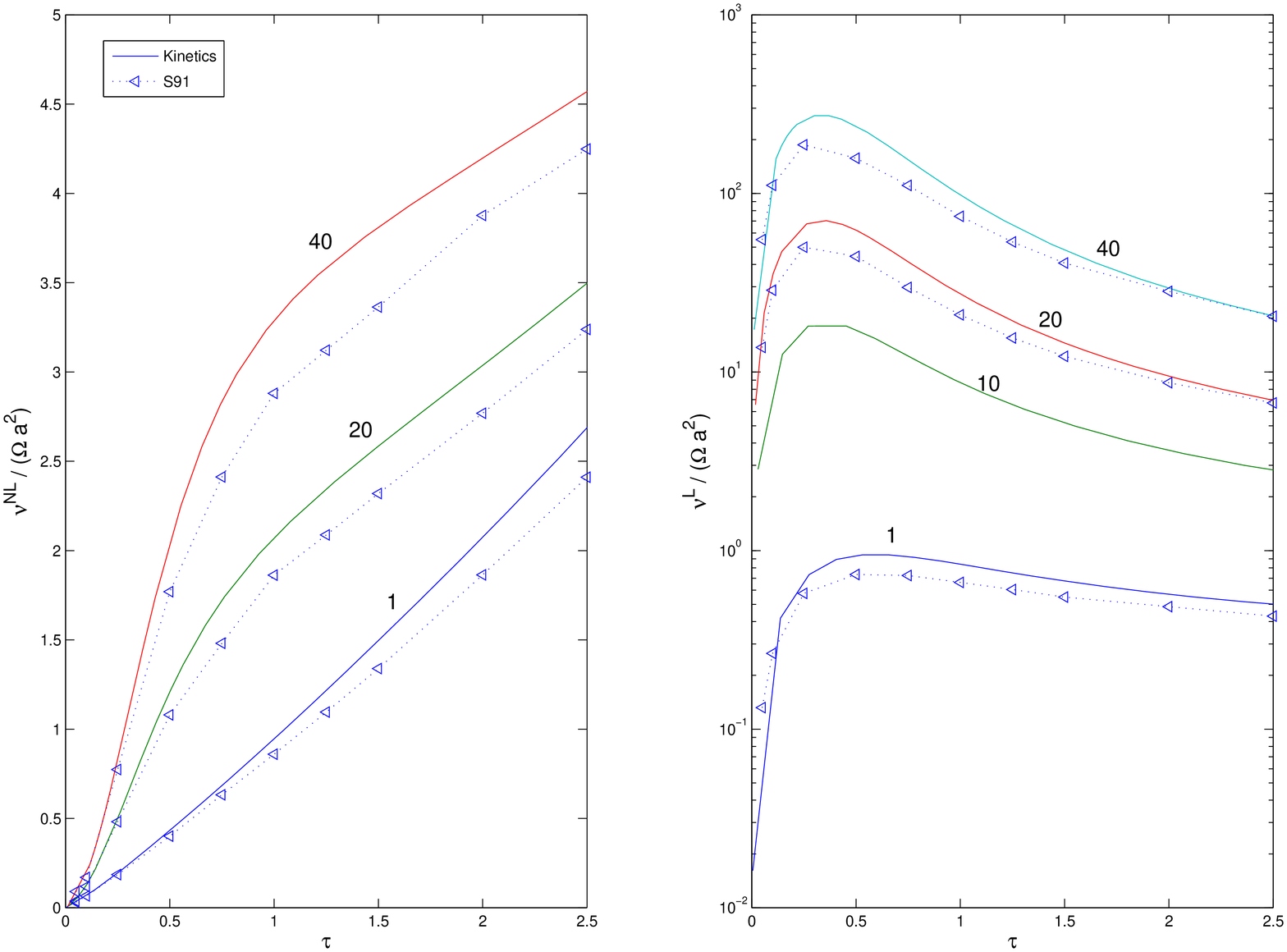}}
\caption{Local and nonlocal components of the kinematic viscosity.
Panel
(a) presents the nonlocal component of the kinematic
  viscosity $\nu^{NL}$, and (b) the local component
  of kinematic viscosity $\nu^{L}$. As earlier different ratios
  $v_c/v_b$ are employed with $p=0.234$, and the solid curves
  represent the polytropic kinetics and the triangles Salo's simulations.
 Both viscosities are scaled by $a^2\Omega$.}
\end{center}
\end{figure}

The magnitude of the velocity dispersion has a direct bearing on the thickness of the disk
as Fig.~3b can attest. Here we can see how the particles'
collisional properties closely constrain important features of the
disk. For the more dilute disks, $H$ is
directly correlated with $c$ and seems to adhere quite well 
to the dilute estimate of $H\approx c/\Omega$. In contrast, it
behaves quite differently in the case of
$v_c/v_b=1$. Here $H$ possesses a subtle turning point at a critical $\tau$
after which it begins to increase.
The increase is caused by the vertical flux of momentum due to collisions,
the
`collisional pressure', which increases with $\tau$.
 As the translational pressure drops and the
disk vertically contracts this contribution will become important and
will work to fatten the disk. If we had included self-gravity, however, the
collisional pressure would be met and overcome and the disk would
flatten further on increasing $\tau$ (see AT86). At some
point, though, this compression must stop.

Fig.~3c is interesting because it shows the disk approach the
dilute energy balance as $v_c$ is increased. The limiting
$\varepsilon(\tau)$ curve in this case being
that of Goldreich and Tremaine (1978) (cf.\ their Fig.~2). In the dense regime (low $v_c$) the
injection of energy by the collisional stress source becomes important and so
each collision must dissipate more energy than otherwise. This
explains the low values of $\varepsilon$ the energy balance requires
when $v_c/v_b=1$ and $0.25$. .

Figs 4a and 4b present the nonlocal and local kinematic viscosity against
$\tau$. Here we find relatively good agreement with Salo (1991) for each
$v_c/v_b$. As with the velocity dispersion,
 the kinetic treatment overestimates the viscosities, especially
 the nonlocal viscosity for larger ratios of $v_c/v_b$. 
Otherwise, the figure displays quite well, in the
relative `convexity' and `concavity' of the curves, the
different behaviours associated with the two regimes of disk --- the warm
and dilute on one hand, and the dense on the other.

 \begin{figure}[!ht]
\begin{center}
\scalebox{.55}{\includegraphics{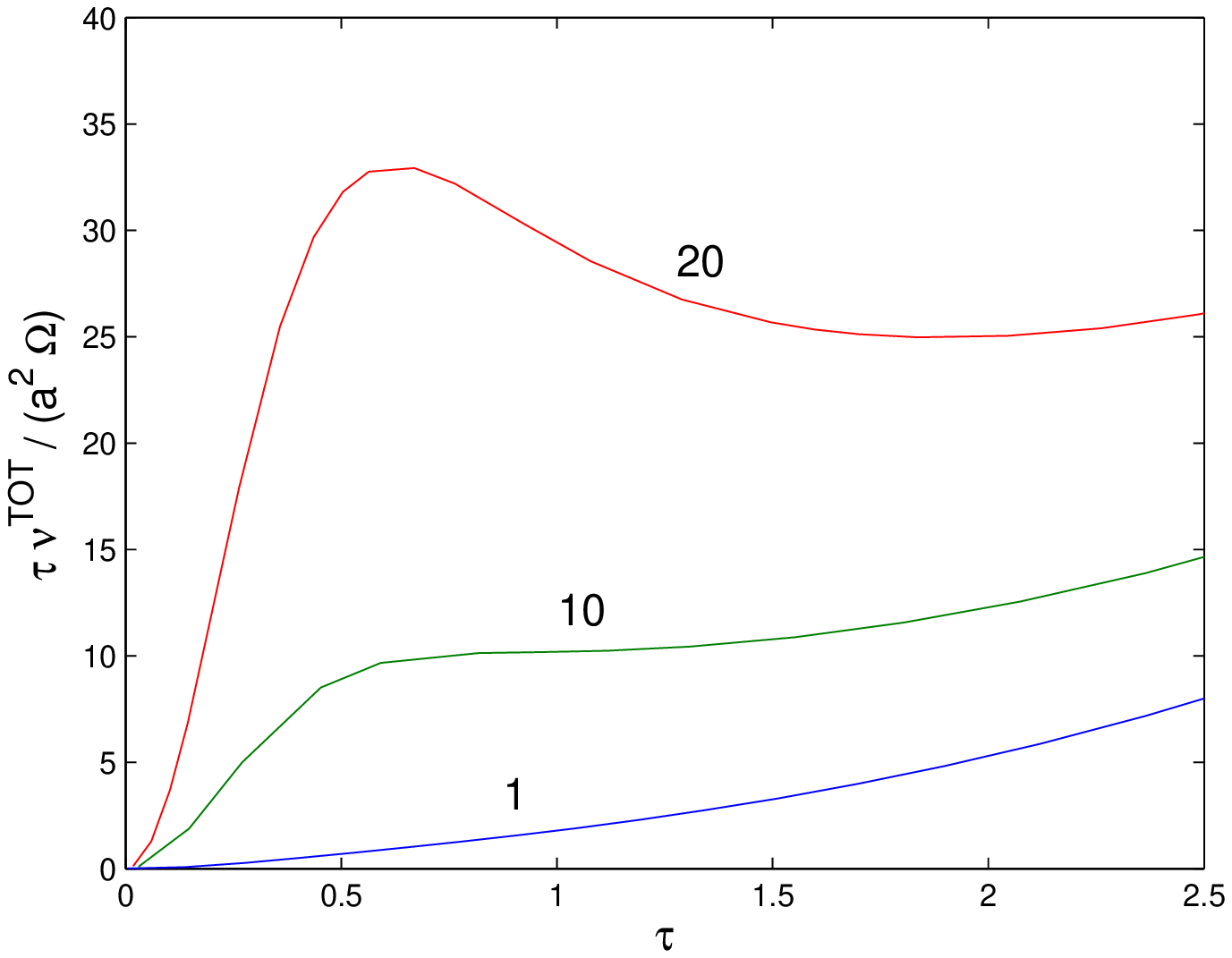}}
\caption{ The total angular momentum flux $\tau \nu=\tau(\nu^{NL}+\nu^{L})$
 versus $\tau$ at different ratios of $v_c/v_b$. The flux is scaled by
  $\Omega a^2$.}
\end{center}
\end{figure}

 \begin{figure}[!ht]
\begin{center}
\scalebox{.475}{\includegraphics{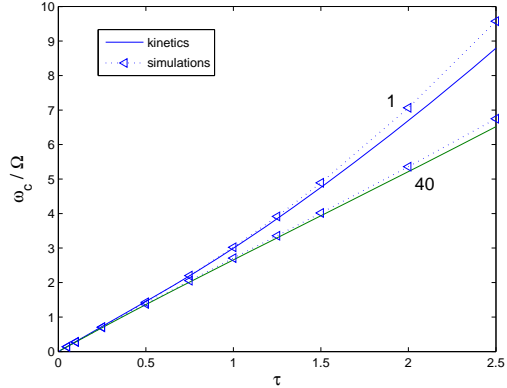}}
\caption{The scaled collision frequency $\omega_c/\Omega$ as a function of optical depth
$\tau$ for two illustrative ratios of $v_c/v_b$.}
\end{center}
\end{figure}

\begin{figure}[!ht]
\begin{center}
\scalebox{.475}{\includegraphics{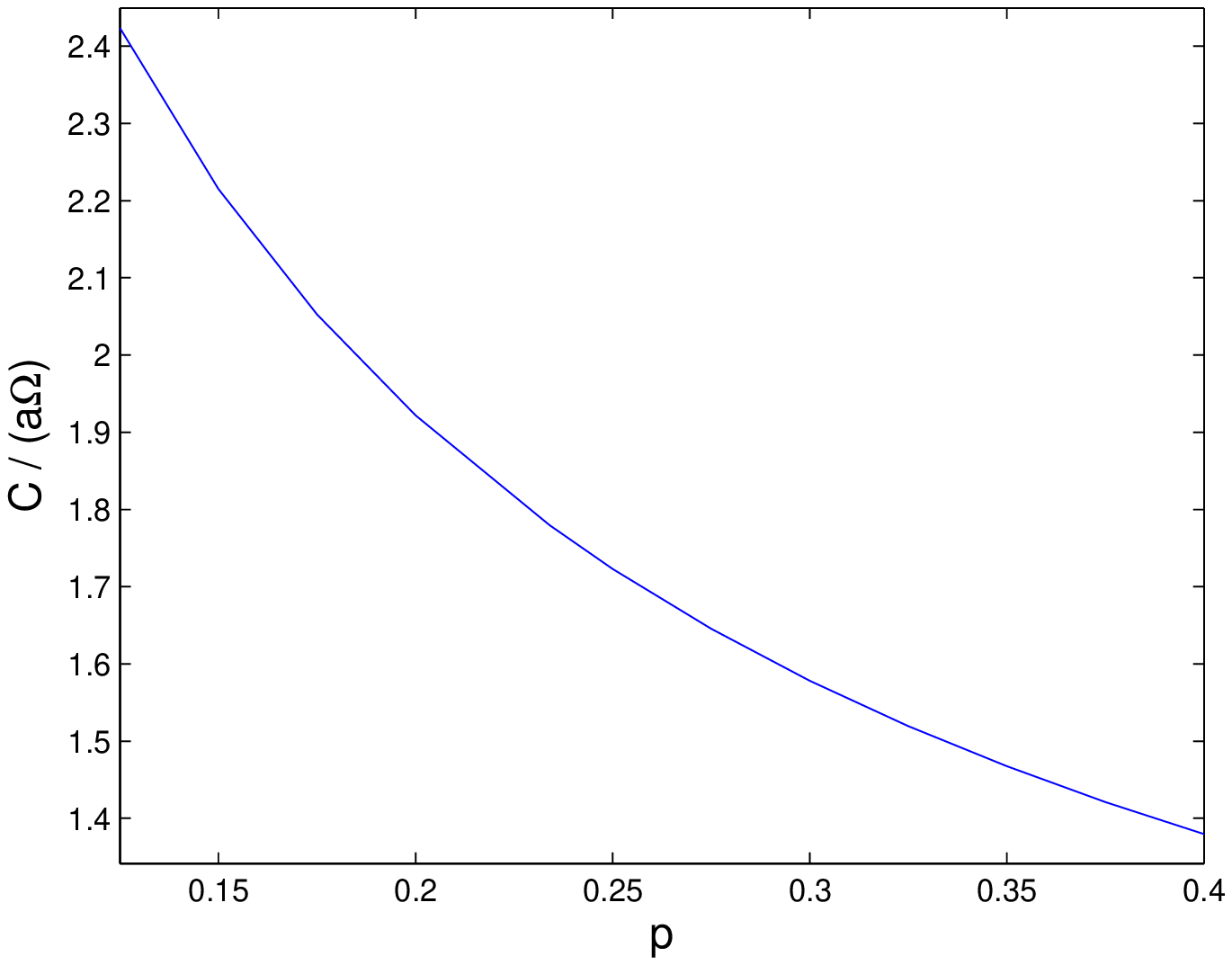}}
\caption{The equilibrium velocity dispersion at $\FF_0=0.2$
  as a function of the exponent $p$ with fixed $v_c=v_b$.}
\end{center}
\end{figure}

In Fig~5 we plot the quantity $\tau \nu$, which is proportional to the angular momentum
 flux and hence a key indicator of the stability of the equilibrium.
 We present three interesting cases. The first, corresponding
to $v_c/v_b=20$, represents a `bimodal' profile; such cases have been discussed by
 H\"{a}meen Antilla (1982)
and  Spahn and Schmidt (2006), and were first revealed in simulations by Salo
 (1991).
 For low to middle
 $\tau$ the disk is in the dilute regime; there the angular momentum flux
 increases sharply and then gently decreases after reaching a turning point.
 As explored in Latter and Ogilvie (2006), the region of large
positive slope does not lead to viscous overstability because
 the stress (dominated by the translational component) is nonlocal in
 time; however, the region of negative slope should give rise to the
 viscous instability. At high optical depths the disk enters the dense regime
and the collision frequency becomes sufficiently large to engender a significant
 collisional stress: the angular momentum flux begins to increase gradually
 under its influence. 

The second curve corresponds to $v_c/v_b=10$, and it shows
 the collisional stress asserting itself at a lower $\tau$, effectively negating the
`window' of decreasing angular momentum flux exhibited by the $v_c/v_b=20$ curve.
 In this intermediate range the
 slope of the curve is near to zero as the opposite tendencies of the local
 and nonlocal stresses cancel. The disk is perhaps marginally stable
 to the viscous instability in this range.

 The third curve
 corresponds to a disk which lies in the dense regime for all $\tau$. Here
 $\nu$ is always dominated by
 the nonlocal component. In this case the angular momentum flux
 gradually increases and may at a critical $\tau$ possess a
gradient sufficiently large to initiate the viscous overstability.

Fig.~6 plots the collision frequencies for two ratios, $v_c/v_b= 40$ and $1$. These illustrate the
dilute regime and the dense regime respectively.
 In the latter case 
$\omega_c$ clearly exhibits its linear dependence on $\tau$, while the
former shows a gentle superlinear dependence for sufficiently large $\tau$,
this mainly arising from the Enskog factor.
The kinetic model uses
expression \eqref{omegac} to evaluate the collision frequency.
 The matching between the two approaches is good in general, though at larger
 $\tau$, there is an increasing discrepancy in the dense case.

Finally, in Fig.~7 we plot the variation of $c$ with respect to the elasticity
exponent $p$ for fixed $v_c=v_b$ and $\FF_0=0.2$. The trend 
reveals that increasing $p$ decreases the equilibrium
velocity dispersion. This follows because an elasticity law with a
larger $p$ varies with impact velocity more steeply.
 With such a law, a
collision characterised by a given impact velocity is more dissipative
than if the slope were more shallow. If collisions are more
dissipative in general then the equilibrium $c$ may be smaller both to minimise
the collision frequency and to maximise the injection of energy by the
collisional stress.

\subsection{Importance of vertical nonlocality}

Having applied the kinetic model to equilibrium calculations, compared
the properties computed with those from simulations, and discussed these
properties more generally, we shall now test some of the assumptions
it uses. First, we examine the approximation of vertical locality,
introduced in Section 2.2.6.  

In order to best test the assumption 
we present the equilibrium characteristics of two representative disks: one without
self-gravity and one with self-gravity. These two cases correspond to
relatively thick and thin disks, respectively, and hence the assumption of
vertical locality will impact on each quite differently. Otherwise, the model
incorporates
\begin{itemize}
\item a preaveraged $\varepsilon$, the piecewise power law, using
  \eqref{whoah!} and \eqref{whoah!2} and with $p=0.234$ and $\textsf{R}_c=2$
\item Gaussian vertical structure.
\end{itemize}
We employ Gaussian vertical structure because this model admits analytical
vertical averages in the vertically nonlocal case, in contrast
to the polytropic model (see Section 3.4). We believe the results we
obtain with the Gaussian profile would also hold qualitatively in the
polytropic case.
Recall that the only change made by the assumption of vertical locality is
 omission of the `nonlocality factor', $\text{exp}\left[-(a^2/H^2)k_z^2\right]$ in the
collision integrals.
 
First we set $G_s=0$ and compute 
non-self-gravitating equilibria with and without vertical
nonlocality. Selected equilibrium properties,
as functions of $\FF_0$, are plotted in Fig.~8. The two models agree closely
 because of the relatively large scale height, $H$,
plotted in Fig.~8b. This quantity is approximately
$2.5$ at its smallest; consequently, the `nonlocality factor'
varies at most between about 1 and $0.85$, and much less at higher central 
filling factors. The discrepancy between the two models can then be at most
 some $15\%$, and it appears its effect is much less.

Next we set $G_s=6.6$, a realistic value for solid ice paticles in the B-ring. The
equilibrium properties are plotted in Fig.~9, as before. The discrepancies
between the two methods are more pronounced, and especially attending
those quantities most closely aligned with the vertical structure, $\tau$ and
$H$. These two deviate substantially at large central filling factors as then
 the disk semithickness approaches a particle radius. For these very thin
disks the `nonlocality factor' can vary between $0.367$ and $1$ and
thus play a significant role. However, we are encouraged by
 the agreement exhibited by the horizontal
properties, such as velocity dispersion, which are not nearly as affected.

 The reason why the vertical
properties are more affected than the horizontal ones is because of the local model's
overestimation of the vertical collisional pressure $P_{zz}^c$ in the height
equation, which is were $H$ and hence $\tau$ are determined effectively.
 In a very thin disk $P_{zz}^c$ will be very small: most colliding particles are in the same plane
and hence nearly all collisional
momentum transport will lie in this plane. But the collisional dynamics of the 
local model is insensitive to
the vertical stratification and hence to
this fact. Evidence for this interpretation can be observed in Fig.~9b
where we see the disk semithickness $H$ increase at a lower $FF_0$ than it
should because of the overestimation of the vertical momentum flux.

\begin{figure}[!ht]
\begin{center}
\scalebox{.65}{\includegraphics{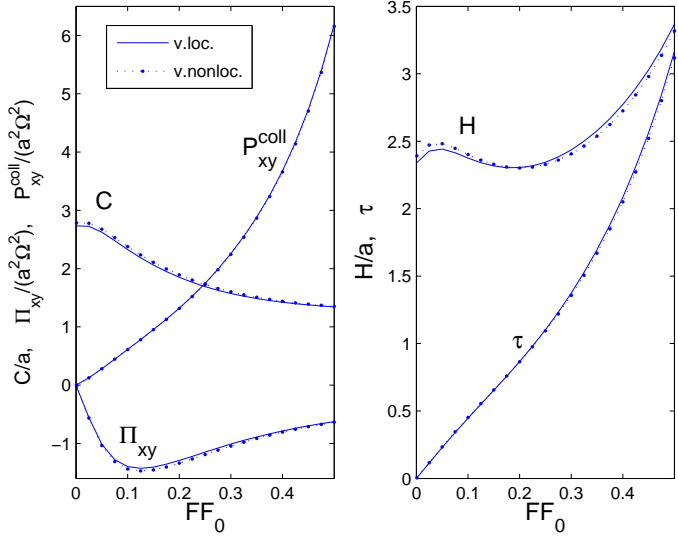}}
\caption{ Comparison between the vertically local model
  (the solid line) against the vertically nonlocal model (triangles)
 in a non-self-gravitating disk, $G_s=0$. In (a) we plot
 selected `horizontal properties' of the equilbria ($C$, $P_{xy}^c$, $\Pi_{xy}$), and in (b)
  the `vertical properties' ($H$, $\tau$).
  These are presented as functions of central filling factor $\FF_0$.
Otherwise we have adopted Gaussian vertical structure and a preaveraged
  $\varepsilon$ with $p=0.234$ and $\textsf{R}_c=2$.}
\end{center}
\end{figure}

\begin{figure}[!ht]
\begin{center}
\scalebox{.65}{\includegraphics{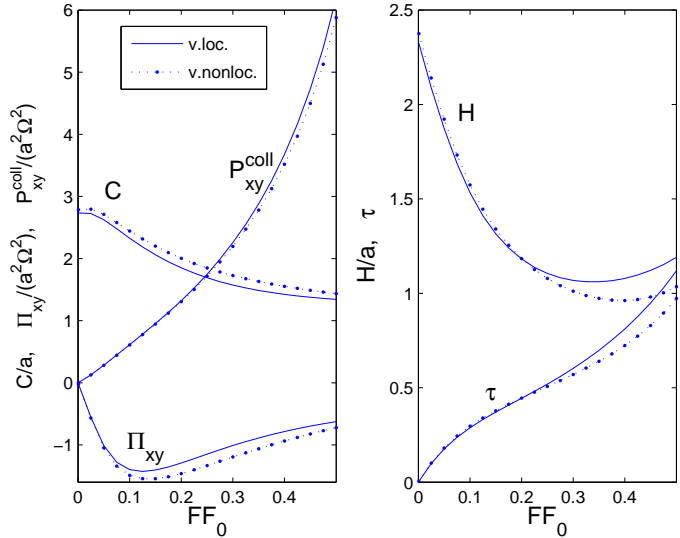}}
\caption{As for Fig.~8 but for a self-gravitating disk, $G_s=6.6$.}
\end{center}
\end{figure}

In summary, however, the assumption of vertical locality is certainly
justified. Very little error is introduced in disks which possess a $H\gtrsim
2a$, and the horizontal properties are approximated adequately 
in self-gravitating disks where the semithickness approaches a
particle radius. It is
only the optical depth $\tau$ and $H$ itself in very thin disks 
that we must be careful
interpreting when $\FF_0$ is high.   
This is pleasing because the assumptiom of vertical locality
 provides a number of mathematical advantages.
 Particularly useful is its decoupling
 of the height equation from the horizontal pressure tensor equations
 and the simplification of the integrands. These simplifications,
 also, are crucial in the polytropic model, and also in the interpolation, and expansion models, as they
 facilitate the complete integration of the collision terms.

\subsection{Validity of preaveraged coefficient of restitution}

\begin{figure}[!ht]
\begin{center}
\scalebox{.85}{\includegraphics{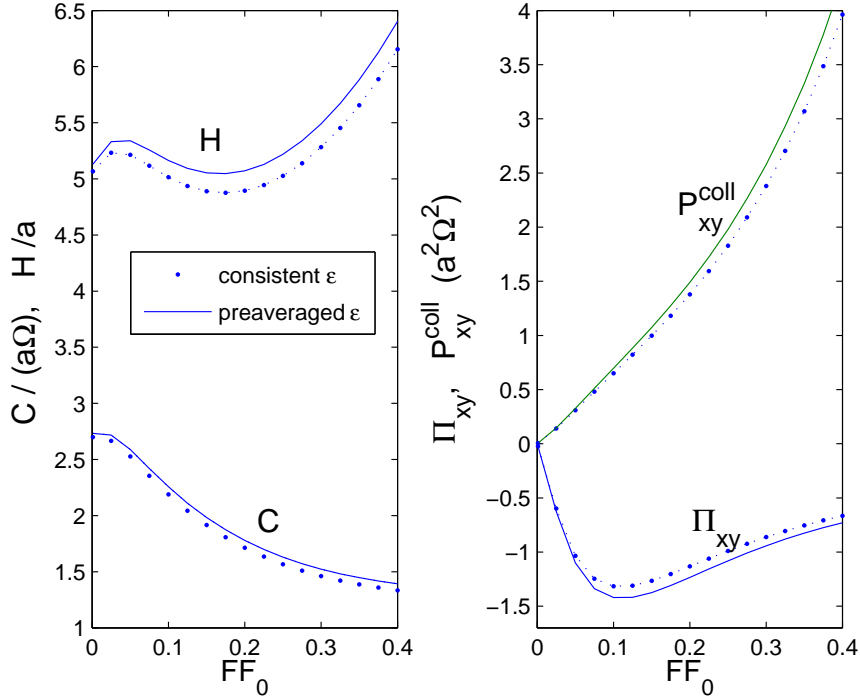}}
\caption{Selected equilibrium properties as computed by a model which
  incorporates a preaveraged $\varepsilon$ and a model that allows
  $\varepsilon$ to vary within the collision integrals. A straight power law was adopted for $\varepsilon$,
  with $p=0.234$ and $\textsf{R}_c=2$. 
The quantities $H$ and $C$ appear in panel (a), and the collisional and
  translational stresses $P_{xy}^c$ and $\Pi_{xy}$
  in panel (b). The vertical structure is assumed to be polytropic.}
\end{center}
\end{figure}

In this section we test the preaveraging approximation against a model which 
incorporates the variation of $\varepsilon$ consistently within the collision
integrals.
 The coefficient of restitution is set as a straight power law: $\varepsilon= (\g\cdot\kk/v_c)^{-p}$ where
$v_c= 0.01$ cm s$^{-1}$ and $p=0.234$.
 This functional form supplies
analytical expressions for both $\langle \varepsilon \rangle_q$ and
$A_\gamma(\varepsilon)$. Expressions for the former are listed in 
\eqref{firstav1} and \eqref{firstav2} and the latter are simple power laws but
with a slightly different `$v_c$' (we omit these expressions). 
 Thus our choice of the straight power law ensures that 
the integrations we need to undertake are only
two-dimensional and not three, as would be the case for the piecewise function.
In addition we set the velocity scale $a\Omega= 0.02$ cm s$^{-1}$ and $G_s=0$.
Otherwise, the
model assumptions are
\begin{itemize}
\item polytropic vertical structure with $\alpha=0.9$
\item vertical locality
\end{itemize}
In Fig.~10 we plot selected equilibrium properties.
 The agreement is quite good generally and excellent for the
collisional stress. The preaveraged model slightly
overestimates the equilibrium velocity dispersion and collisional pressure 
tensor. 
Overall, however, these results reveal that this approximation gives
excellent qualitative and good quantitative agreement.

\subsection{Interpolation model}

Finally, we investigate the interpolation model introduced in Section
2.2.8 and worked out in detail in Appendix A. The $F$ functions were
interpolated on the range $\xi\in [-0.75,0.75]$ with a mini-max method
which returned a relative error of at most 3\%. The $F_{ij}$ which
appear in Eqs \eqref{interpen} were
computed as
\begin{align*}
&F_{11}= 1.9619, \qquad F_{12}= 5.3174 , \qquad F_{13}= 6.545, \qquad
F_{14}= 3.5450, \\
&\qquad\qquad F_{21}= 1.0000, \qquad F_{22}= 1.7725, \qquad F_{23}= 0.9352, \\
&\qquad\qquad F_{31}= 1.7725, \qquad F_{32}= 4.5401, \qquad F_{33}= 3.5450
\end{align*}
(to 4 decimal places). Also we assume
\begin{itemize}
\item polytropic vertical structure with $\alpha=0.9$,
\item vertical locality,
\item a piecewise power law, which is preaveraged with $p=0.234$ and $\textsf{R}_c=2$
\end{itemize}
Because the collision integrals are linear in $\ani$ 
the equilibrium problem is greatly simplified. We can solve Eqs
\eqref{eqm1}-\eqref{eqm2} analytically for the
$xx$, $xy$ and $yy$ components of $\ani$, leaving a single nonlinear
algebraic equation for $c$ in terms of the governing parameters,
$\tau$, $\mathsf{R}_c$, $p$, $G_s$. The equilibrium calculation is then
 rapidly completed by a one-dimensional Newton-Raphson algorithm.

A comparison of the equilibrium properties as computed by the
interpolation and exact method are presented in Fig.~11. 
The agreement here is good throughout the full range of $\tau$ which is
somewhat surprising: we expected discrepancies near $\tau=0$ arising 
from the truncation in anisotropy ($\ani$). Perhaps this accounts for
 the deviation of disk semithickness $H$ at low $\tau$,
but it is puzzling
 that this vertical property is singled out. In general, 
given the truncation in $\ani$, the excellent agreement
for general $\tau$ suggests that dense rings exhibit a level of 
anisotropy which can be captured adequately by a linear model.

\begin{figure}
\begin{center}
\scalebox{.75}{\includegraphics{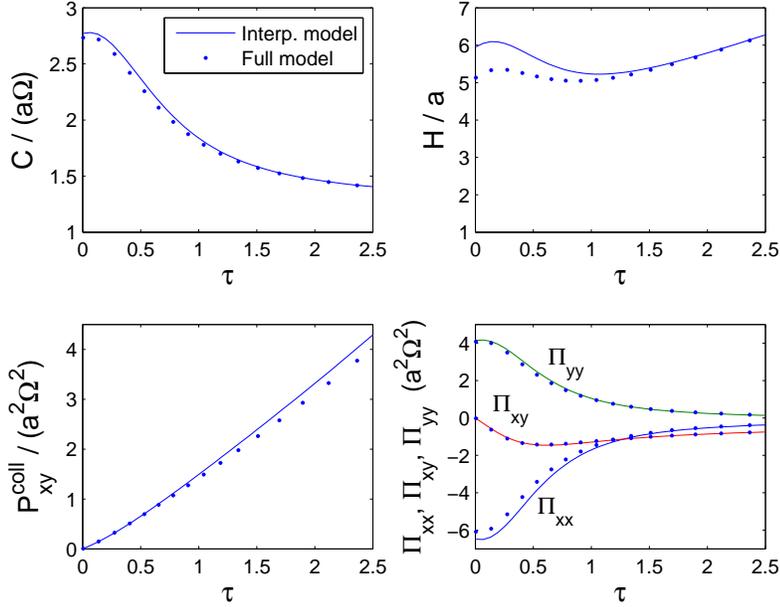}}
\caption{Selected equilibrium properties as functions of $\tau$ computed by the
  full model (points)
  and the interpolation model (solid line). Both models adopt polytropic
  vertical structure ($\alpha=0.9$) and a preaveraged
  $\varepsilon$ with $p=0.234$ and
  $\textsf{R}_c=2$. The velocity dispersion $C$ is plotted in panel (a), the
  semithickness $H$ in (b), the collisional stress $P_{xy}^c$ in (c), and
  components of the translational stress $\mathbf{\Pi}$ in (d)}
\end{center}
\end{figure}

Because this method reduces the collision integrals to algebraic
expressions (albeit complicated), more advanced linear and nonlinear
analyses of the ring kinetics are easier to undertake. Moreover, its success in
equilibrium calculations encourages its application in these investigations.

\section{Linear stability}

Now that the statistics of the dense disk equilibrium have been
established, we shall examine its response to small perturbations. Our
interest is principally in the growth rates and stability of the 
 overstable modes as functions of the kinetic parameters. We will not
supply an exhaustive survey in this paper but rather illustrate the principal
features while also touching on a comparison with the simulations of Salo
\emph{et al.} (2001).

\subsection{Linearised equations}

The disk admits steady
homogeneous solutions characterised by the
constant optical depth $\tau_0$ and velocity dispersion tensor
$\mathbf{W}^0$. Suppose that we perturb these
equilibria with an infinitesimal axisymmetric disturbance
\begin{align*}
\tau &=\tau_0+ \tau'(x,t),\\
 \u &= -(3\Omega\,x/2)\mathbf{1}_y + \{ u(x,t)\mathbf{1}_x + v(x,t)\mathbf{1}_y\}, \\
\mathbf{W} &= \mathbf{W}^0 + \mathbf{W}'(x,t), \\
\Phi_D &= \Phi_D^0+\Phi_D'(x,t),
\end{align*}
The system is subsequently nondimensionalised along the following lines:
 time with respect to $\Omega^{-1}$, space with
respect to $C_0/\Omega$, velocity with $C_0$, and velocity dispersion
tensor with $C_0^2$. We now use the nondimensionalised quantities but
with no change in notation.

The linearised equations read
\begin{align}
\d_t\tau' &= -\tau_0\,\d_x\,u, \\
\d_t u &= -\d_x \Phi'+2v-\d_x\left[W_{xx}'+(P^\text{c}_{xx}/N)'\right]-\frac{1}{\tau_0}(W_{xx}+P_{xx}^{\text{c}}/N)_0\,\d_x\tau',
\label{momeq} \\
\d_t v &= -\frac{1}{2}\,u
-\d_x\left[W_{xy}'+(P^\text{c}_{xy}/N)'\right]-\frac{1}{\tau_0}\,(W_{xy}+P_{xy}^{\text{c}}/N)_0\,\d_x\tau',
\\
\d_t W_{xx}'&= -2W_{xx}^0\d_x u + 4W_{xy}'+ (Q_{xx}/N)', \\
\d_t W_{xy}' &= -W_{xy}^0 \d_x u - W_{xx}^0 \d_x v
+2W_{yy}'-\frac{1}{2}W_{xx}'+(Q_{xy}/N)', \\
\d_t W_{yy}' &= -2W_{xy}^0 \d_x v - W_{xy}' + (Q_{yy}/N)', \\
\d_t W_{zz}' &= (Q_{zz}/N)', 
\end{align}
where
the perturbed collision terms, $Q_{ij}'$ and $(P^\text{c}_{ij})'$,
are expressed in the form
\begin{align*}
Q_{ij}'&= \left(\frac{\d Q_{ij} }{\d \tau }\right)_0 \tau'
  + \left(\frac{\d Q_{ij} }{\d W_{xx} }\right)_0 W_{xx}' 
  + \left(\frac{\d Q_{ij} }{\d  W_{xy}}\right)_0 W_{xy}' \\
&  +\left(\frac{\d Q_{ij}}{\d W_{yy}}\right)_0 W_{yy}'
+\left(\frac{\d Q_{ij}}{\d W_{zz}}\right)_0 W_{zz}' 
 +\left(\frac{\d Q_{ij}}{\d e_{xx}}\right)_0 e_{xx}'
+\left(\frac{\d Q_{ij}}{\d e_{xy}}\right)_0 e_{xy}',
\end{align*}
and $e_{ij}$ is the rate of strain tensor. 
To these we add the perturbed Poisson equation, in order to obtain $\Phi'$,
and the
perturbed height equation, in order to relate the perturbed central filling
factor $\FF_0'$ with $\tau'$.

Once we express the radial perturbation as Fourier modes $\propto
\text{exp}(ikx+st)$,
we let the nondimensionalised solution of the linearised Poisson equation take the form
\begin{equation} \label{phi}
\Phi_D'=-\frac{2g}{|k|}\tau'
\end{equation}
 (Binney and
Tremaine, 1987), where
\begin{equation}\label{gprob}
 g= \left(\frac{\,m\,G}{a^3
    \Omega}\right)\left(\frac{a\Omega}{C_0}\right)=G_s\,R,
\end{equation}
and is equal to $(Q\tau_0)^{-1}$, where $Q$ is the Toomre parameter.
Once this is substituted we obtain a seventh order
 algebraic eigenvalue problem for the growth rate $s$. This in turn
 supplies a seventh order dispersion relation of the same structure as
 Eq.~(25) in Latter and Ogilvie (2006), but in which the coefficients are
  functions of $G_s$, $g$, $k$, the equilibrium values of
 $\mathbf{W}$ and $\mathbf{P}^c$, and the derivatives of
$\mathbf{W}$ and $\mathbf{P}^c$. 
If we adopt the general dense formalism
devised in Section 2.2.6, evaluating these coefficients requires
the numerical calculation of a number of surface integrals and their
derivatives.

Before continuing, we will say a few words about the parameter $g$ given by
Eq.~\eqref{gprob}. 
Though it appears that $g$ simply derives from the quantities
$G_s$ and $R$, a quick calculation shows that it can take large values
in a typical dense equilibrium, and hence $Q$ can take very low values.
 For example, if $G_s=6.6$, $a=1$ m, $\tau\approx 1$ and $C/(a\Omega)\approx 1.5$ (see
Fig.~9), we compute $Q \approx 0.23 $. Kinetic models characterised by these low values
of $Q$ exhibit viscous overstability on arbitrarily short scales. Some aspects
of this issue have been examined in Latter and Ogilvie (2006) in the context
of viscous instability.
Part of the problem originates in the small velocity dispersion of axisymmetric 
self-gravitating equilibria, an admittedly artificial scenario. A
 real self-gravitating disk will be subject to non-axisymmetric instability
 for a $Q$ less than about 2, and the recurrent wake structures it engenders
 will heat the disk sufficiently to keep $Q$ above unity. In addition,
 gravitational scattering and focusing (effects we neglect) will supply extra
 heat sources. Self-gravitating simulations bear out these point and show that
 $Q$ is typically unity or higher (Salo, 1992a).
 The second important issue is our omission of the
third order moments which are important on short scales and which should
extinguish instability. These considerations will not be addressed in this paper,
but we will revisit it in a future nonlinear axisymmetric study where
a solution to this problem is more urgent.

\subsection{Results}

The model we employ incorporates 
\begin{itemize}
\item polytropic vertical structure with $\alpha=0.9$,
\item vertical locality, and
\item a preaveraged piecewise
power law for $\varepsilon$.
\end{itemize}
 Both the equilibrium and its linear stability 
 depend on six parameters: $\tau_0$, $p$, $a$, $m$, $\Omega$, and $v_c$,
 all of which can be determined by
observation or experiment.
The nondimensionalised system is parametrised by the dimensionless quantities $\tau$,
$p$, $\textsf{R}_c$ and $G_s$. 

The perturbed collision terms depends
 on the perturbations of $\FF_0$ via the Enskog factor. But this must
be related to the other perturbations before we can proceed. The relevant
 equation can be obtained via \eqref{HHH2}: 
\begin{align*}
\FF_0'&=\frac{1}{H_0 + \FF_0 (\d H/\d \FF)_0}\,
\left\{
\frac{4\,\Gamma(\alpha+\tfrac{3}{2})}{3\sqrt{\pi}\,\Gamma(\alpha+1)}\,\tau'-
\FF_0\left[ \left(\frac{\d H }{\d W_{xx} }\right)_0 W_{xx}' \right.\right.\\
& \hskip1.5cm + \left(\frac{\d H }{\d  W_{xy}}\right)_0 W_{xy}' 
  +\left(\frac{\d H}{\d W_{yy}}\right)_0 W_{yy}' \\
&\hskip2cm\left.\left.  +\left(\frac{\d H}{\d W_{zz}}\right)_0 W_{zz}' 
 +\left(\frac{\d H}{\d e_{xx}}\right)_0 e_{xx}'
+\left(\frac{\d H}{\d e_{xy}}\right)_0 e_{xy}'\right] \right\}.
\end{align*}
This equation, which relates the perturbations of
$\FF_0$, $H$ and $\tau$, conveys information pertaining to vertical disk structure into
the linear dynamics via the equilibrium $H$ and its derivatives.
 In particular, it
introduces
 vertical self-gravity into the problem.
 Self-gravity also
appears
 in the $x$ component of the
momentum equation, \eqref{momeq}.

Once the linearised equations are sorted, our procedure is to step
through $\FF_0$, evaluate the equilibrium at each point and obtain the
dispersion relation, which is then solved for various $k$ (or wavelength).
 We perform three runs: 
\begin{itemize}
\item  without self-gravity in either the height
equation or in the momentum equation, Fig.\ 12;
\item with self-gravity in the height equation alone (thereby modelling
  only its enhancement of the
  vertical restoring forces), Fig.\ 13;
\item with self-gravity in both equations, Fig.\ 14.
\end{itemize}
The algorithm evaluates the growth rates of the seven linear modes
discussed in Latter and Ogilvie (2006).
 These are: the two potential viscous overstable modes, the potential viscous
instability mode, the energy mode, the two potential anisotropic
overstable modes, and the relaxation mode.
We, however, concentrate on the two potentially
overstable ones (left and right-going waves). 
First, their curves of marginal stability are plotted in the
$(\tau,\lambda)$ plane, where $\lambda=2\pi/k$ is wavelength. Then, to these
we add
the curves of constant growth or decay, which we truncate as they
collapse onto the line of marginal stability.

We adopt the Bridges \emph{et al.} (1984) elasticity law, so as to
compare with the simulations of Salo \emph{et al.} (2001); thus $p=0.234$ and $v_c=0.0077$ cm
s$^{-1}$. The particles are assumed to possess radii of 100 cm and the
shearing sheet is located at a radius of $100,000$ km.  As a result $a\Omega=
0.0195$ cm s$^{-1}$ and $\textsf{R}_c= a\,\Omega/v_c= 2.53$.
 We vary $G_s$ so as to
compare our results with the simulations, in which the
particles' internal density took several values less than $900$ kg m$^{-3}$
(solid ice). 
 In addition, Salo \emph{et al.} conducted a number of
simulations where the vertical component of 
self-gravity was mimicked by an enhancement of the particles' vertical
epicyclic frequency, $\Omega_z$. The effective $G_s$ associated with such as
enhancement canot be readily determined, though we have the estimate
\begin{equation} \label{omegaz}
\Omega_z^2 = \Omega^2 + (\d_z^2\phi_D)_{z=0} \approx \Omega^2\,(1+ 3 G_s\, \FF_0),
\end{equation}
in which it has been assumed that $\phi_D$ varies much slower with $x$ than with $z$. 
Typically, Salo \emph{et al.}'s simulations set $\Omega_z/\Omega=3.6$
which corresponds to very large $G_s$, especially for low $\FF_0$. When $\FF_0$
is about 0.3, $G_s$ is still greater than 10. This is surely an overestimate,
and Eq.~\eqref{omegaz} should be treated with some caution.

\subsubsection{Growth rates}

We find that a non-self-gravitating disk is stable to the viscous
instability for all the optical depths we examine: $0<\tau<5$. This 
conforms with the dense gas $N$-body simulations.
Above a critical
$\tau$ the viscous overstability occurs. This value is approximately $\tau=2.5$ which corresponds to a central 
filling factor of $0.375$ (cf. Fig.~12). In contrast, the $N$-body simulations of Salo find
 overstability only for very high optical depths, of order $4$. 
This disagreement is not unexpected: high filling
factor disks in monodiperse simulations can `relax' their close packing via the formation of
layers or `sheets'. Such disks, as a consequence, can possess
filling factors below the critical level of overstability,
despite being quite optically thick.

 The stability curves of the long modes differ
very little from each other: once the critical $\tau$ is reached nearly all the overstable
modes become unstable, which mirrors the hydrodynamical stability curves of
Schmidt \emph{et al.} (2001). As the disk makes the transition to instability,
modes on lengths of some 200 m will grow fastest. The shortest overstable
scales are approximately $80$---$90$ m. These are a little longer than usually
predicted by hydrodynamics and simulations.

\begin{figure}[!ht]
\begin{center}
\scalebox{.8}{\includegraphics{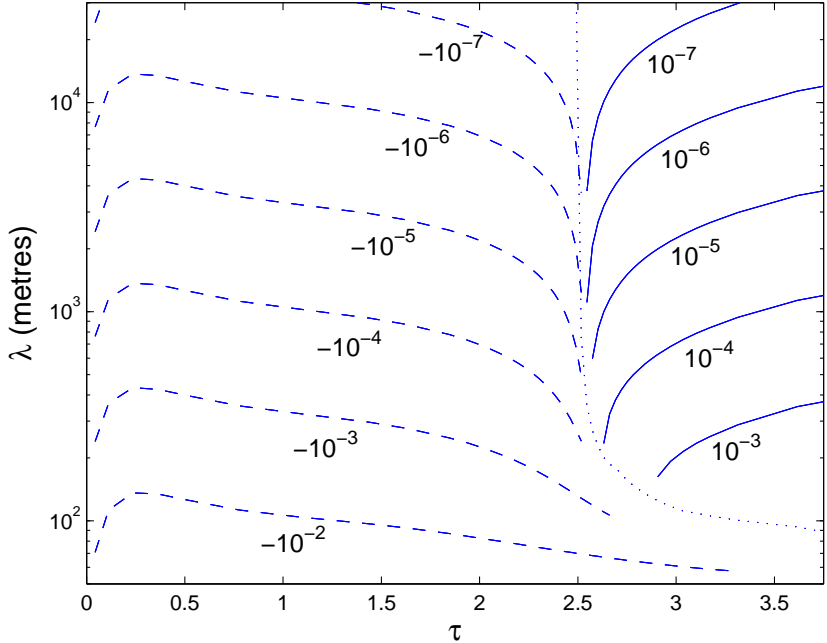}}
\caption{The curves of constant growth and decay rates, Re($s$), for the overstable
  modes in the $(\tau,\lambda)$ plane in a non-self-gravitating disk ($G_s=0$,
  $g=0$).
The model adopts polytropic vertical structure and a pre-averaged $\varepsilon$ with $p=0.234$
  and $\textsf{R}_c=2.53$.
 Negative growth
  rates are plotted with dashed lines and positive growth rates with
  solid lines.
 The curve of marginal stability, $\text{Re}(s)=0$, is the
  dotted line.}
\end{center}
\end{figure}

\begin{figure}[!ht]
\begin{center}
\scalebox{.8}{\includegraphics{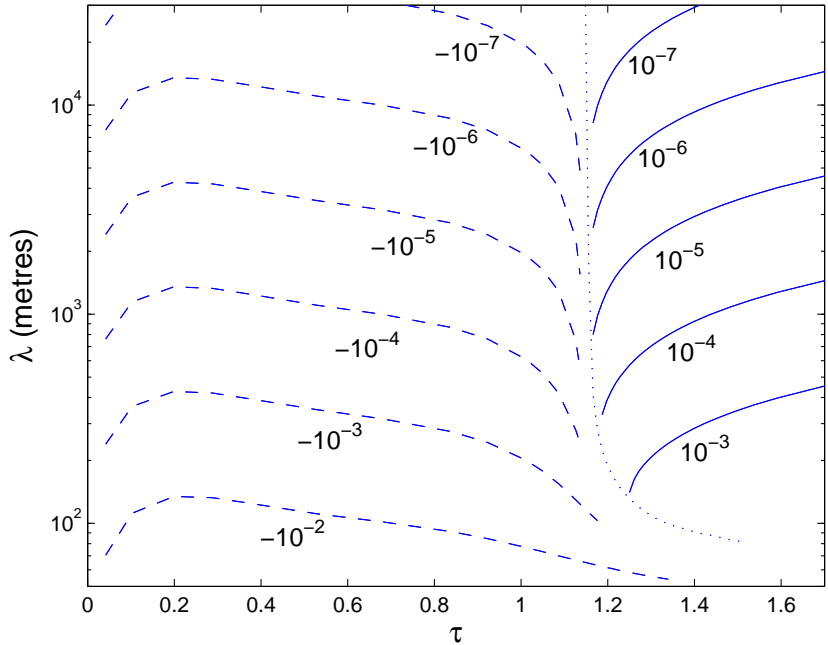}}
\caption{The curves of constant growth and decay rates Re($s$) for the overstable
  modes in the $(\tau,\lambda)$ plane, in a disk where self-gravity is
  included in the height equation but not elsewhere. We have set $G_s=3$ and $g=0$.}
\end{center}
\end{figure}

\vskip0.4cm

In order to examine the contribution of the vertical component of self-gravity
we set $G_s=3.0$ in the height equation, but zero in the momentum
equation. This captures the vertical compression caused by self-gravity 
(and the subsequent increase in collision frequency and
its rate of change)
but will not alter the structure of the dispersion relation. The value
adopted for $G_s$ corresponds to internal particle densities of $\rho_p=
408\,\text{kg}\,\text{m}^{-3}$, which is less than solid ice but is
illustrative nonetheless.

The curves of constant growth and decay are plotted in
Fig.~13. These are similar to those in Fig.~12; the chief difference lies in
a shift to lower optical depths and slightly shorter scales.
 The critical $\tau$ in this case is
$\approx 1.15$ and the critical $\FF_0$ approximately $0.335$. 

The fastest growing modes at the stability transition are now on scales of about
150---200 m, and the shortest affected scales are about 60---70 m. These agree
a little closer with the generic hydrodynamic profiles.

\begin{figure}[!ht]
\begin{center}
\scalebox{.8}{\includegraphics{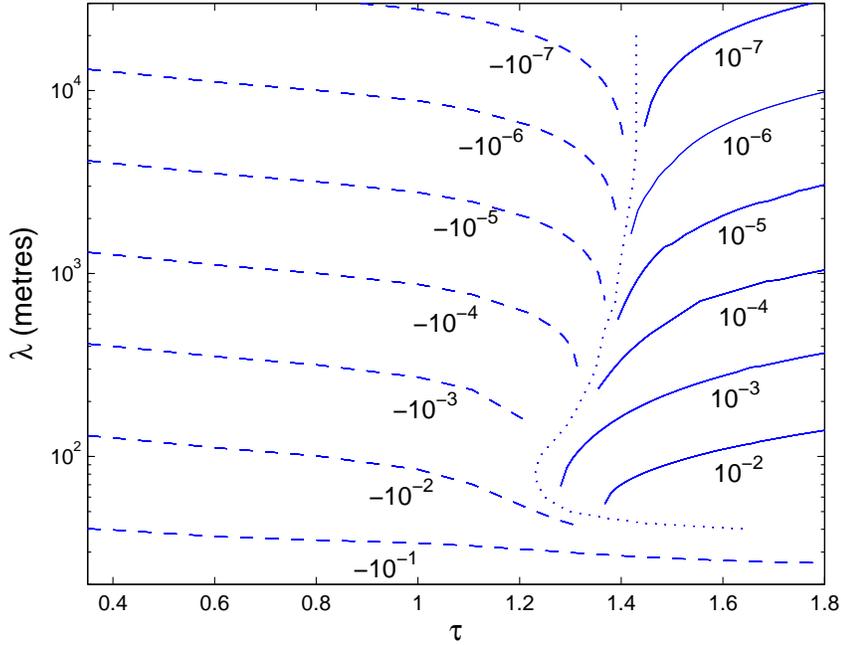}}
\caption{The curves of constant growth rate Re($s$) for the overstable
  modes in the $(\tau,\lambda)$ plane, in the case when self-gravity
  is properly included in both height and momentum equations. Here $G_s=1.65$
  and $g= G_s R$, cf. Eq.~\eqref{gprob}.}
\end{center}
\end{figure}

\vskip0.4cm
Next we include self-gravity in both the height
equation and in the $x$ component of the momentum equation. We seek to compare
directly with the self-gravitating run of Salo \emph{et al.} in which
$\rho_p=225$ kg m$^{-3}$. Therefore, $G_s= 1.65$ by Eq.~\eqref{Gss} and $g$ takes
the appropriate value from Eq.~\eqref{gprob}.
 The subsequent curves of
constant growth and decay are plotted in Fig.~14.

 With the addition of self-gravity in the momentum equation it is the intermediate
lengthscales ($\sim 100$ m) which become unstable first, as illustrated by the
pronounced salient in the curve of marginal stability.
The longest wavelengths are unaffected.
On the other hand, decreasing $G_s$ from $3$ to $1.65$ increases the critical $\tau$ 
of overstability on the longest scales, from about 1.15 to approximately
1.43 (which corresponds to a central filling factor of $0.34$). In contrast, the lowest critical
$\tau$ for overstability on the intermediate scales is about 1.23 (for
$\lambda\approx 100$ m).  

These results are in good agreement with the simulations, which find that
overstability begins at approximately $1.2$ (see Fig~5a in Salo \emph{et al.},
2001) and on scales of 100---150 m. This conformity contrasts with the poor
agreement in the non-self-gravitating run and proceeds essentially from the
introduction of the vertical self-gravity.
 In simulations the self-gravity negates, to some extent, the
relaxation of packing fraction and allows the higher filling factors
necessary for overstability. It thus removes the obstacle that prevented the
kinetic model and the simulations agreeing. 

Finally we set $G_s=6.6$, corresponding to solid ice particles,
 $\rho_p=900\,\text{kg}\,\text{m}^{-3}$.
 The parameter
$g$ was set to its appropriate value from \eqref{gprob}, and thus the Toomre
parameter was generally very small. The critical $\tau$ is now shifted to
 about 0.8 for long scales and somewhat less for intermediate scales.
 As mentioned earlier, in Section 4.1, we
find that the overstability is not quenched on the short scales, thus the `bulge'
in the stability curve associated with self-gravity (cf. Fig.~14) extends down
to lengths of a particle radius. This behaviour is surely unphysical and
derives not only from the small $Q$ but also the omission of the third order
moments in the pressure tensor equation.

\subsubsection{Marginal curves}

Finally we let the elastic properties of the particles vary by letting
$p$ and the parameter $\textsf{R}_c=
a\Omega/v_c$ take other values. Marginal curves of Re($s)=0$ for viscous
overstability are computed in $(p,\tau)$ space for given $k$ and
$\textsf{R}_c$ for a self-gravitating disk with $G_s=6.6$ but with $g=0$.
 These curves are obtained from a simple (but robust) 
bisection root-finding method. Fig.~15 plots three illustrative curves 
for a mode with
$k=0.01$, which corresponds roughly to $\lambda= 600$ m. The parameter $\textsf{R}_c$ takes the values $2.53$,
$0.67$, and $0.25$. The first two correspond to 100 cm size
particles at a radius of $100,000$ km
in the B-ring with the elasticity data of Bridges \emph{et al.} (1984) and
Supulver \emph{et al.} (1995) respectively.

\begin{figure}[!ht]
\begin{center}
\scalebox{.7}{\includegraphics{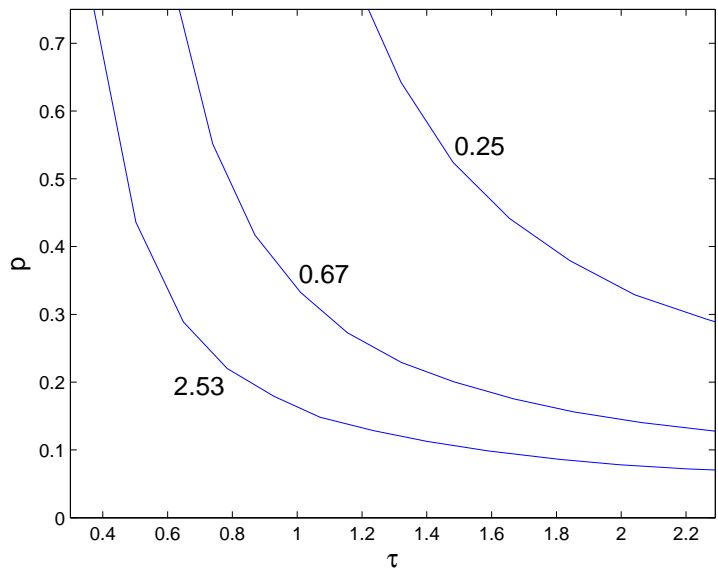}}
\caption{Marginal curves of overstability in the $(p,\tau)$ plane for
  $k=0.01$, $G_s=6.6$ and various $\textsf{R}_c=a\Omega/v_c$. Regions above the curves are
  overstable, regions below are stable.}
\end{center}
\end{figure}

The stability trend here is easy to interpret: the larger $\tau$, $p$, and $\textsf{R}_c$
the more vulnerable the disk to viscous overstability.
 The last correlation derives
from the fact that the greater $\textsf{R}_c$ the more important the background shear
flow, and consequently the greater the collisional viscous stress. As $\textsf{R}_c\to 0$
the nonlocal contributions vanish. The effect of $p$ lies in its
control of the equilibrium disk velocity dispersion. The larger $p$
the lower $C$, and hence the more important the collisional viscous
stress (cf.~Fig.~7). 

Another fact to note is the sensitivity of the critical $\tau$ to the
elasticity laws. The Bridges \emph{et al.} data (for which $p=0.234$)
 gives a critical $\tau$ of
roughly 0.8, but the Supulver \emph{et al.}
data (for which $p=0.19$) gives a value nearer to 1.5. In contrast, in these two cases and
more generally,
the critical $\FF_0$ does not change as dramatically
and lies somewhere in the interval 0.32---0.375. We conclude that
 the viscous overstability responds essentially to a sufficiently large filling factor,
 the critical value of which does not vary significantly with respect to the
 collision law (or 
self-gravity). It rather than $\tau$ is the key parameter governing the
onset of instability.

Though the stability results presented in Fig.~15 hold only for a mode of
length some $600$ m, the stability
of all the scales should follow qualitatively the same trend, irrespective of
the precise value of $G_s$ and $g$. Indeed these results
 are consistent with the self-gravitating simulations of Salo \emph{et
 al.} (2001)
 when they vary $\textsf{R}_c$ (cf. their Fig.~5b).

Given that optical depth is a key observational property, and that it depends
so delicately on the elastic properties of the particles, our detailed model
 is a useful first step in constraining the collisional
parameters of ring particles. Certainly the model needs refining,
particularly with respect to self-gravity, but it does provide an account of dynamical
processes which can be compared directly with observations. Studies
such as Salo and Karjalainen's (2003) combined dynamical and photometric
modelling have already made inroads into such comparisons.

\section{Conclusion}

In summary, we have developed a workable formalism with which to
attack the dynamics of a dense ring, one in which the collisional transport/production
 processes and filling factor effects are fully
included,
 as is the dependence of
$\varepsilon$ on impact velocity. The model both simplifies and
generalises the work of AT86 and as a consequence
encourages us to conduct more advanced dynamical analyses.

 The theory is founded on
that of Enskog kinetics and thus shares its assumptions regarding the
neglect of ternary or higher collisions and the modelling of space and velocity
correlations. An additional assumption is the limiting of
the velocity dispersion's spatial variation to scales significantly
longer than a particle radius (i.e. vertical isothermality). Also the
velocity distribution is assumed to be a triaxial Gaussian, and the vertical
structure of the number density to take a predetermined form, either a
Gaussian or a polytrope. The theory must be supplemented, with the functional forms of the Enskog
factor and the coefficient of restitution. 

The full model can be further simplified if additional 
assumptions are made, such as those of locality in density, a preaveraged
coefficient of restitution, or if the collision integrands are
interpolated and anisotropy assumed small. Each of
these models fared well in comparisons with the exact approach (Sections 3.9
--- 3.11). Thus we employed them generally.

We undertook a comparison of equilibria computed by the kinetics of Araki and
Tremaine (1986)
and the simulations of Wisdom and Tremaine (1988) and Salo (1991).
 The former two comparisons exposed
the shortcomings of the Gaussian model of vertical structure at moderate and
high central filling factors in monodisperse disks, and revealed an
error in the numerical implementation of the Araki and Tremaine model. 
It was also shown that a polytropic model of vertical structure performed
well in these regimes and acceptably at lower filling factors.

The polytropic kinetic model was tested in equilibrium calculations against the
results of Salo's (1991) $N$-body simulations. Overall the agreement
was good, with the major discrepancies within the range of error introduced by
 the various aproximations.

Next we probed the linear theory of the kinetic model and reproduced qualitatively
much of the behaviour presented by the particle simulations of Salo \emph{et
  al.} (2001). We find that without self-gravity a disk composed of 100 cm radius
particles
 exhibits
viscous overstability at a critical optical depth of about 2.5. This is a
significantly lower value than Salo \emph{et al.}'s, which is approximately 4,
and proceeds essentially from the kinetic model's insistence on a fixed
vertical structure. A simulation can `relax' its close packing by the
 formation of vertical sheets and thus remain below the critical filling factor
necessary to instigate overstability. (Note however that the sheets are artefacts
of a monodisperse model and hence unrealistic.)  
This discrepancy vanishes when
self-gravity is properly included and agreement  between theory and
simulations is quite good for comparatively light particles. In this case
modes of 100 m are the most unstable and possess a critical optical depth of
about 1.23.
In general, inclusion of the vertical component of self-gravity decreases the
critical optical depth above which overstability manifests, while inclusion of
the radial component of self-gravity induces overstability preferentially on
intermediate scales, while leaving the longest scales untouched.

When the self-gravity is large, for instance when particles are as dense as
solid ice, the linear dynamics on short scales is incorrectly described. The
viscous overstability then extends to lengths of about a particle radius, which
is unphysical. This arises because of the omission of third order moments from
the pressure tensor equations and also because of the very low velocity dispersions
which characterise axisymmetric disks in which
gravitational collisions have been neglected. In reality,
non-axisymmetric wakes, and gravitational collisions, warm the disk sufficiently
and on short scales `thermal diffusion' extinguishes instability.
 This issue must be addressed more fully in
 future nonlinear studies which employ the
kinetic model and assume axisymmetry.

\vskip0.3cm

 Within the model assumptions,
we predict that overstability can occur at any radius of Saturn's
rings provided the optical depth exceeds a critical
value. This value depends rather sensitively on the
parameters of the collision law, in our case $p$ and $v_c$; in the
B-ring, a plausible range of critical $\tau$ is between 0.8 and 1.6 (cf.\ Fig.~15). 
 It must be said, however, that these estimates omit the effects of the
gravitational wakes and size distribution both of which militate against
the overstability's onset. Even so, 
the dense B-ring seems the most likely place in
 which overstability occurs, as it exhibits unambiguously high optical
 thicknesses.
 Finescale structure in such optically thick regions we
 conclude should be associated with this instability.

\vskip0.3cm

Our kinetic model could be developed in a number of ways.
For instance, it would be straightforward, though tedious, to
include the spin degrees of freedom and, possibly, size distribution. In
 the former case, we may use similar techniques to simplify the
spin collision integrals. If a similarly workable formalism was
 developed
 we could examine, amongst a
 great many things, the action of particle spin on the onset of the
 overstability. Simulations have already begun probing this problem and await
 a kinetic comparison (Morishima and Salo, 2006).
However, it should be noted that at present the
 frictional properties of ring particles are poorly understood.
Also, the effect of gravitational encounters between particles should also be
included. This is an important but difficult task. At present the model omits 
this important heating mechanism and thus underestimates the velocity
dispersion of self-gravitating equilibria.

There are a numerous ring-related problems on which the full or
interpolation formalism could
be put to use immediately.
 The most obvious is the
nonlinear saturation of the viscous overstability. The seven evolutionary
equations could be numerically simulated to investigate the long-term
evolution of the instability and how the elasticity parameters
influence the statistics of the quasi-steady state to which it leads.
The damping
of nonlinear density waves launched by moons such as Prometheus, Pandora, etc
 could also be studied, as well
as the wavy edges and wakes induced by Daphnis and Pan in the Keeler
and Encke gaps. Theoretical models and their predictions
 have suffered from an approximate description of the viscosity in
 the ring. The full dense gas model could supply a more correct
 account of these effects.
These would be formidable computational problems as at each time step
and each spatial grid point we would need to evaluate a number of
two-dimensional integrals. Such projects may best be tackled with the
interpolation model with its algebraic expressions for the collision terms. 
That said continuum models, such as those we develop here, can reach the
longer length and time scales necessary to fully describe the nonlinear
behaviour. $N$-body methods are still too computationally intensive,
particularly when self-gravity is added, to feasibly capture these scales.

\vskip4cm

\textbf{Acknowledgements}
\vskip0.3cm
We are extremely grateful to the referees Juergen Schmidt and Heikki Salo for
their thorough checking of the manuscript. We also thanks them for the
computational results they made available to us in the process of
revision. HNL acknowledges funding from the Cambridge Commonwealth Trust,
Trinity College Cambridge, and STFC.

\section{Appendix: Interpolated collision terms}
Presented here are the collision terms computed from the interpolation
method to order $\ani$. 
We have:
\begin{align}
\mathbf{q}= \frac{2}{\pi}\,a^2(1+\varepsilon) \widetilde{Y}\left[ (1+\varepsilon)\mathbf{L}_1- 2 \mathbf{L}_2 \right]
\end{align}
where the $\mathbf{L}_1$ tensor is defined by
\begin{align*}
\mathbf{L}_1&= 4\pi c^3 \left\{ F_{10}\left(\tfrac{1}{3}\mathbf{I}_2 + \tfrac{3}{10} \mathbf{I}_4:\ani \right)
  - \frac{a}{c}\, F_{11} \left( \tfrac{1}{5} \mathbf{I}_4:\mathbf{e} + 
                           \tfrac{1}{7}\mathbf{I}_6:\ani:\mathbf{e}\right) \right. \\
& \hskip1cm \left. + \frac{a^2}{c^2}\, F_{12} \left(\tfrac{1}{7} \mathbf{I}_6:\mathbf{e}^2 + 
                                    \tfrac{1}{18} \mathbf{I}_8:\ani:\mathbf{e}^2\right) 
- \frac{a^3}{c^3}\, F_{13}\left(\tfrac{1}{9} \mathbf{I}_8:\mathbf{e}^3\right) \right\}.
\end{align*}
and the $\mathbf{L}_2$ tensor by
\begin{align*}
\mathbf{L}_2 &= 4\pi c^3 \left\{ \tfrac{2}{3}F_{30}\left( \mathbf{I}_2 + \tfrac{6}{5} \ani\right) 
                      - \frac{2a}{15c} F_{31}\left( \ani\mathbf{e}+\mathbf{e}\ani + 2\mathbf{e}+
                     \text{Tr}(\mathbf{e})\ani +\text{Tr}(\mathbf{e})\mathbf{I}_2\right) \right. \\ 
   & \hskip1cm \frac{a^2}{105c^2} F_{32} \left( 8[\mathbf{e}^2\ani +  \ani\mathbf{e}^2]
                + 4 \text{Tr}(\mathbf{e})[\mathbf{e}\ani + \ani\mathbf{e}]
              +  2[ \text{Tr}(\mathbf{e})^2 + 2\text{Tr}(\mathbf{e}^2)]\ani \right. \\  
   &\left. \hskip1cm \left.  +30\mathbf{I}_6:\mathbf{e}^2- \tfrac{35}{3}\mathbf{I}_8:\ani:\mathbf{e}^2\right)\right\}.
\end{align*}
The $F_{ij}$ are pure numbers which are determined from the interpolation method and interpolation range;
 $\mathbf{I}_2$ is the rank 2 identity tensor and the other tensors are defined through
\begin{align*}
&\mathbf{I}_4:\ani = \tfrac{2}{3}\ani, \\
&\mathbf{I}_4:\mathbf{e} = \tfrac{2}{3}\mathbf{e}+\tfrac{1}{3}\text{Tr}(\mathbf{e})\,\mathbf{I}_2, \\
&\mathbf{I}_6:\ani:\mathbf{e} = \tfrac{2}{15}\left\{ 2(\ani\mathbf{e} + \mathbf{e}\ani)
                + \text{Tr}(\mathbf{e})\ani + \text{Tr}(\mathbf{e}\ani)\mathbf{I}_2 \right\}, \\
&\mathbf{I}_6:\mathbf{e}^2 = \tfrac{1}{15}\left\{ 8\mathbf{e}^2 + 4\text{Tr}(\mathbf{e})\mathbf{e}
               +\text{Tr}(\mathbf{e})^2\mathbf{I}_2 + 2 \text{Tr}(\mathbf{e}^2)\mathbf{I}_2 \right\}, \\ 
&\mathbf{I}_8:\ani:\mathbf{e}^2 = \tfrac{2}{105} \left\{ 
                8(\ani\mathbf{e}^2 + \mathbf{e}\ani\mathbf{e}+\mathbf{e}^2\ani) 
               +4\text{Tr}(\mathbf{e})( \ani\mathbf{e}+ \mathbf{e}\ani) \right. \\
& \hskip2cm \left.    +4 \text{Tr}(\ani\mathbf{e})\mathbf{e} + 
              [\text{Tr}(\mathbf{e})^2 + 2 \text{Tr}(\mathbf{e}^2)]\ani 
               + 2 \text{Tr}(\mathbf{e})\text{Tr}(\ani\mathbf{e})\mathbf{I}_2 
              + 4 \text{Tr}(\mathbf{e}\ani\mathbf{e})\mathbf{I}_2 \right\} \\
&\mathbf{I}_8:\mathbf{e}^3 = \tfrac{1}{105}\left\{
              48\mathbf{e}^3+ 24 \text{Tr}(\mathbf{e})\mathbf{e}^2 + 
                6[\text{Tr}(\mathbf{e})^2+2\text{Tr}(\mathbf{e}^2)]\mathbf{e} \right. \\
 & \hskip5cm  \left.   +[\text{Tr}(\mathbf{e})^3+ 8\text{Tr}(\mathbf{e^3})+ 
                6\text{Tr}(\mathbf{e})\text{Tr}(\mathbf{e}^2)] \mathbf{I}_2 \right\}.
\end{align*}

The collision momentum flux tensor is
\begin{align*}
\PP&= 8\sqrt{\pi}(1+\varepsilon)\widetilde{Y}\,a^3\,c^2\left\{ 
             F_{20} \left( \tfrac{1}{3}\mathbf{I}_2 +\tfrac{1}{5}\mathbf{I}_4:\ani \right) \right. \\
& \hskip2cm \left.  
-\frac{a}{c}\, F_{21} \left( \tfrac{1}{5}\mathbf{I}_4:\mathbf{e}+\tfrac{1}{14}\mathbf{I}_6:\ani:\mathbf{e}\right)
+ \frac{a^2}{7c^2}\, F_{22}\, \mathbf{I}_6:\mathbf{e}^2 \right\}.
\end{align*}

If the $F_{jn}$ are set to zero except for $F_{10}=2$, $F_{20}=\sqrt{\pi}$, and
$F_{30}=1$ then we recover the dilute gas formalism of Goldreich and
Tremaine (1978) to order $\ani$. See Section 2.2.9.

\end{document}